\documentclass{aa}

\usepackage{graphicx}
\usepackage{txfonts}
\usepackage{amstext}
\usepackage{xcolor}
\usepackage{color}
\usepackage{threeparttable}
\usepackage{overpic}
\usepackage{multirow}
\usepackage[normalem]{ulem} 
\usepackage{amsmath} 
\usepackage{hyperref}
\hypersetup{
  colorlinks   = true, 
  urlcolor     = blue, 
  linkcolor    = blue, 
  citecolor   = blue 
}

\makeatletter
\let\jnl@style=\rm
\def\ref@jnl#1{{\jnl@style#1}}

\def\aj{\ref@jnl{AJ}}                   
\def\actaa{\ref@jnl{Acta Astron.}}      
\def\araa{\ref@jnl{ARA\&A}}             
\def\apj{\ref@jnl{ApJ}}                 
\def\apjl{\ref@jnl{ApJ}}                
\def\apjs{\ref@jnl{ApJS}}               
\def\ao{\ref@jnl{Appl.~Opt.}}           
\def\apss{\ref@jnl{Ap\&SS}}             
\def\aap{\ref@jnl{A\&A}}                
\def\aapr{\ref@jnl{A\&A~Rev.}}          
\def\aaps{\ref@jnl{A\&AS}}              
\def\azh{\ref@jnl{AZh}}                 
\def\baas{\ref@jnl{BAAS}}               
\def\bac{\ref@jnl{Bull. astr. Inst. Czechosl.}}
\def\caa{\ref@jnl{Chinese Astron. Astrophys.}}
\def\cjaa{\ref@jnl{Chinese J. Astron. Astrophys.}}
\def\icarus{\ref@jnl{Icarus}}           
\def\jcap{\ref@jnl{J. Cosmology Astropart. Phys.}}
\def\jrasc{\ref@jnl{JRASC}}             
\def\memras{\ref@jnl{MmRAS}}            
\def\mnras{\ref@jnl{MNRAS}}             
\def\na{\ref@jnl{New A}}                
\def\nar{\ref@jnl{New A Rev.}}          
\def\pra{\ref@jnl{Phys.~Rev.~A}}        
\def\prb{\ref@jnl{Phys.~Rev.~B}}        
\def\prc{\ref@jnl{Phys.~Rev.~C}}        
\def\prd{\ref@jnl{Phys.~Rev.~D}}        
\def\pre{\ref@jnl{Phys.~Rev.~E}}        
\def\prl{\ref@jnl{Phys.~Rev.~Lett.}}    
\def\pasa{\ref@jnl{PASA}}               
\def\pasp{\ref@jnl{PASP}}               
\def\pasj{\ref@jnl{PASJ}}               
\def\rmxaa{\ref@jnl{Rev. Mexicana Astron. Astrofis.}}%
\def\qjras{\ref@jnl{QJRAS}}             
\def\skytel{\ref@jnl{S\&T}}             
\def\solphys{\ref@jnl{Sol.~Phys.}}      
\def\sovast{\ref@jnl{Soviet~Ast.}}      
\def\ssr{\ref@jnl{Space~Sci.~Rev.}}     
\def\zap{\ref@jnl{ZAp}}                 
\def\nat{\ref@jnl{Nature}}              
\def\iaucirc{\ref@jnl{IAU~Circ.}}       
\def\aplett{\ref@jnl{Astrophys.~Lett.}} 
\def\apspr{\ref@jnl{Astrophys.~Space~Phys.~Res.}}
\def\bain{\ref@jnl{Bull.~Astron.~Inst.~Netherlands}} 
\def\fcp{\ref@jnl{Fund.~Cosmic~Phys.}}  
\def\gca{\ref@jnl{Geochim.~Cosmochim.~Acta}}   
\def\grl{\ref@jnl{Geophys.~Res.~Lett.}} 
\def\jcp{\ref@jnl{J.~Chem.~Phys.}}      
\def\jgr{\ref@jnl{J.~Geophys.~Res.}}    
\def\jqsrt{\ref@jnl{J.~Quant.~Spec.~Radiat.~Transf.}}
\def\memsai{\ref@jnl{Mem.~Soc.~Astron.~Italiana}}
\def\nphysa{\ref@jnl{Nucl.~Phys.~A}}   
\def\physrep{\ref@jnl{Phys.~Rep.}}   
\def\physscr{\ref@jnl{Phys.~Scr}}   
\def\planss{\ref@jnl{Planet.~Space~Sci.}}   
\def\procspie{\ref@jnl{Proc.~SPIE}}   

\makeatother

\defcitealias{2016MNRAS.460.2526O}{OH16}

\begin{document}

\title{The period-gap cataclysmic variable CzeV404~Her: \\ 
       A link between SW~Sex and SU~UMa systems
\thanks{Based on observations obtained at San Pedro Mart\'ir Observatory, UNAM, Baja California, Mexico, Ond\v{r}ejov Observatory, Czech Republic, 
Observatorio del Roquede los Muchachos, Spain,
BS Observatory, Zl\'{\i}n, Czech Republic, 
and La Silla Observatory, Chile.
Table \ref{table_OC} is available in electronic form at {\tt www.aanda.org}. }}

\titlerunning{The period-gap cataclysmic variable CzeV404~Her}{}
\authorrunning{J. K\'{a}ra, et al. }
\author{J. K\'{a}ra \inst{1}
\and S. Zharikov\inst{2}
\and M. Wolf\inst{1} 
\and H. Ku\v{c}\'{a}kov\'{a}\inst{1,3,4,5}
\and P. Caga\v{s}\inst{5,6} 
\and A. L. Medina Rodriguez\inst{2}
\and M. Ma\v{s}ek\inst{5,7}
}
\institute{Astronomical Institute, Faculty of Mathematics and Physics, Charles University, V~Hole\v{s}ovi\v{c}k\'ach~2, CZ-180~00~Praha~8, \\
            Czech Republic, \email{honza.kara.7@gmail.com}
 \and Universidad Nacional Aut\'{o}noma de M\'{e}xico, Instituto de Astronom\'{i}a, AP 106,  Ensenada, 22800, BC, M\'{e}xico 
\and Astronomical Institute, Academy of Sciences, Fri\v{c}ova 298, 251 65 Ond\v{r}ejov, Czech Republic
\and Research Centre for Theoretical Physics and Astrophysics, Institute of Physics, Silesian University in Opava, Bezru\v{c}ovo n\'{a}m. 13,
746 01 Opava, Czech Republic
\and Variable Star and Exoplanet Section, Czech Astronomical Society, Fri\v{c}ova 298, 251 65 Ond\v{r}ejov
\and BSObservatory, Modr\'{a} 587, CZ-760~01~Zl\'{\i}n, Czech Republic
\and FZU – Institute of Physics of the Czech Academy of Sciences, Na Slovance 1999/2, Prague 182 21, Czech Republic}
\date{Received \today}
        
\abstract{We present a new study of the eclipsing cataclysmic variable CzeV404~Her ($P_{\rm orb}=0.098$ d) that is located in the period gap.}{This report determines the origin of the object and the system parameters and probes the accretion flow structure of the system.}{We conducted simultaneous time-resolved photometric and spectroscopic observations of CzeV404~Her. We applied our light-curve modelling techniques and the Doppler tomography method to determine the system parameters and analyse the structure of the accretion disk.}
{We found that the system has a massive white dwarf $M_{\rm WD}=1.00(2)$ M$_{\sun}$, a mass ratio of $q=0.16$, and a relatively hot secondary with an effective temperature $T_2 = 4100(50)$~K. The system inclination is $i=78.8^\circ$. The accretion disk spreads out to the tidal limitation radius and has an extended hot spot or line region. The hot spot or line is hotter than the remaining outer part of the disk in quiescence or in intermediate state, but does not stand out completely from the disk flux in (super)outbursts.}
{We claim that this object represents a link between two distinct classes of SU~UMa-type and SW~Sex-type cataclysmic variables. The accretion flow structure in the disk corresponds to the SW~Sex systems, but the physical conditions inside the disk fit the behaviour of SU~UMa-type objects.}


\maketitle
\section{Introduction}  

Cataclysmic variables (CVs) are semi-detached binaries consisting of a white dwarf as a primary and a late-type (K-M) secondary star or a brown dwarf \citep{1995CAS....28.....W}. The secondary overfills its Roche lobe 
and transfers matter onto the primary. The orbital period of CVs ranges from $\sim 80$ min to  $\sim
10$~hours. If the primary has a weak magnetic field ($ < 10^{5} \text{ G}$), the transferred matter forms an
accretion disk around the white dwarf.
In some CVs, called dwarf novae (DNe), the accretion disk can
fluctuate between a low- and a high-temperature or density state. This explosive transition has an outbursting behaviour. The amplitudes of DNe outbursts range between 2 -- 8 magnitudes and last for 2 -- 20 days.

The range of orbital periods between $\sim$2.15~and $\sim$3.18~hours is called the period gap. A
significant deficiency of CVs is observed there \citep{2011ApJS..194...28K}. It is assumed that the lack of CVs 
inside the period gap is caused by the secondary star becoming fully convective. This causes the magnetic breaking to shut off, whereby long-period CVs lose angular momentum  and sustain mass transfer
between their components \citep{2016MNRAS.457.3867Z,2018ApJ...868...60G}. Unable to lose its angular momentum, 
the binary evolves towards shorter orbital periods as a detached binary. When the system reaches an orbital
period of $\sim 2$ hours, mass transfer restarts. For shorter periods until the period minimum, the 
emission of gravitational waves becomes the leading process of losing the angular momentum, 
and this is efficient enough for the mass transfer to continue.

The small number of CVs in the period gap \citep{2003Ap.....46..114K} includes eclipsing systems
\citep{2012MmSAI..83..614D}, which provide an opportunity to determine system parameters 
and to probe the evolutionary theories. These systems include polars, DNes, novas,  and nova-likes. 
In recent years, the list has been slightly extended mostly by the newly discovered polars   MLS110213:022733+130617
\citep{2015MNRAS.451.4183S}, CRTS J035010.7+323230  \citep{2019MNRAS.488.2881M},  2PBC J0658.0-1746
\citep{2019MNRAS.489.1044B}, DNes  CzeV404~Her \citep{2014IBVS.6097....1C}, and RXS~J003828.7+250920 \citep{2016Ap.....59..304P}.

\begin{figure}[t]
        \centering
        \includegraphics[width=0.48\textwidth]{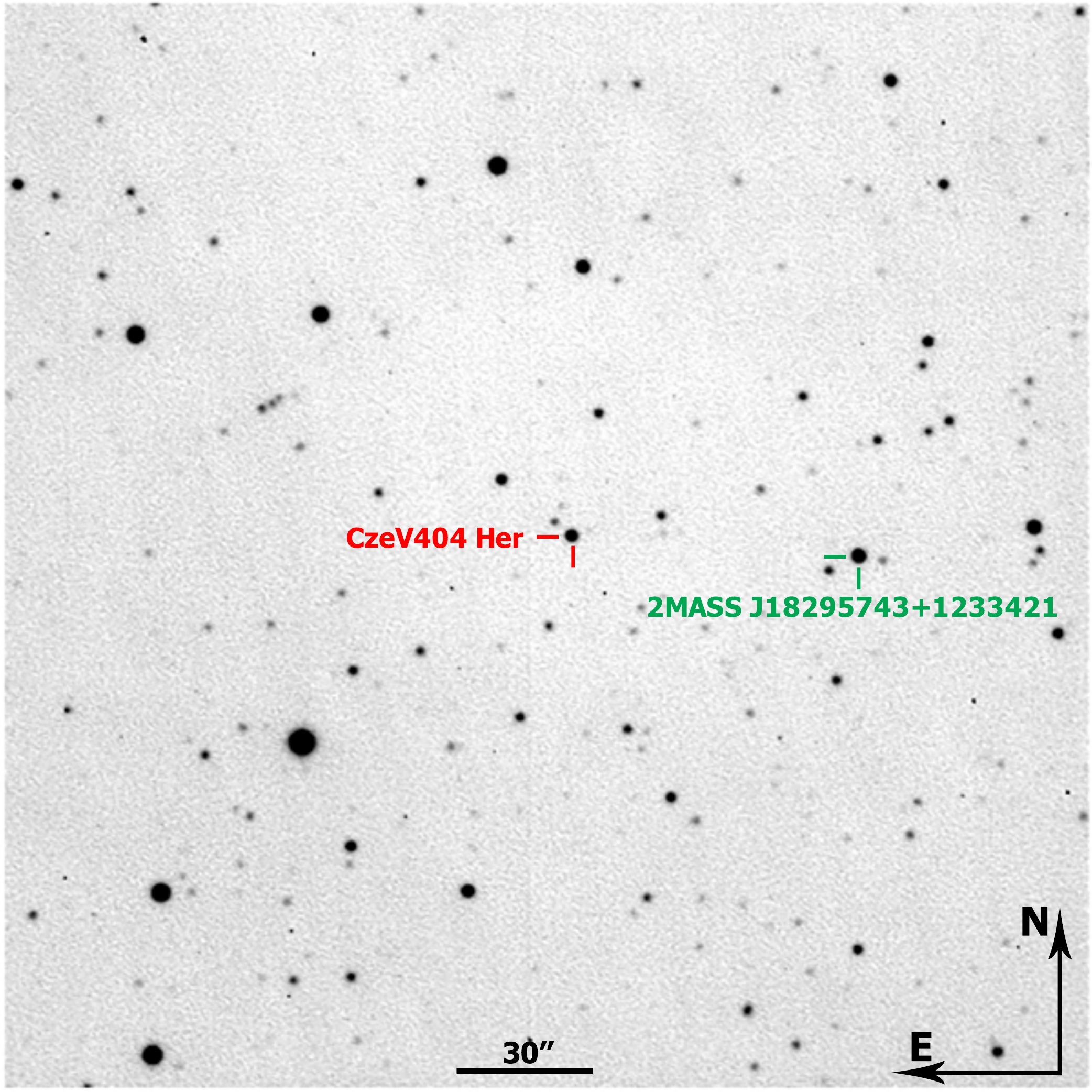}
        \caption{Finding chart showing the position of CzeV404~Her (red) and the selected comparison star (green) in a frame obtained at La Silla observatory with the Danish 1.54m telescope. The field of view is $4 \times 4$ arcminutes.}
        \label{Fig:01}
\end{figure}

\begin{figure*}[t]
        \centering
        \includegraphics[width=0.98\textwidth]{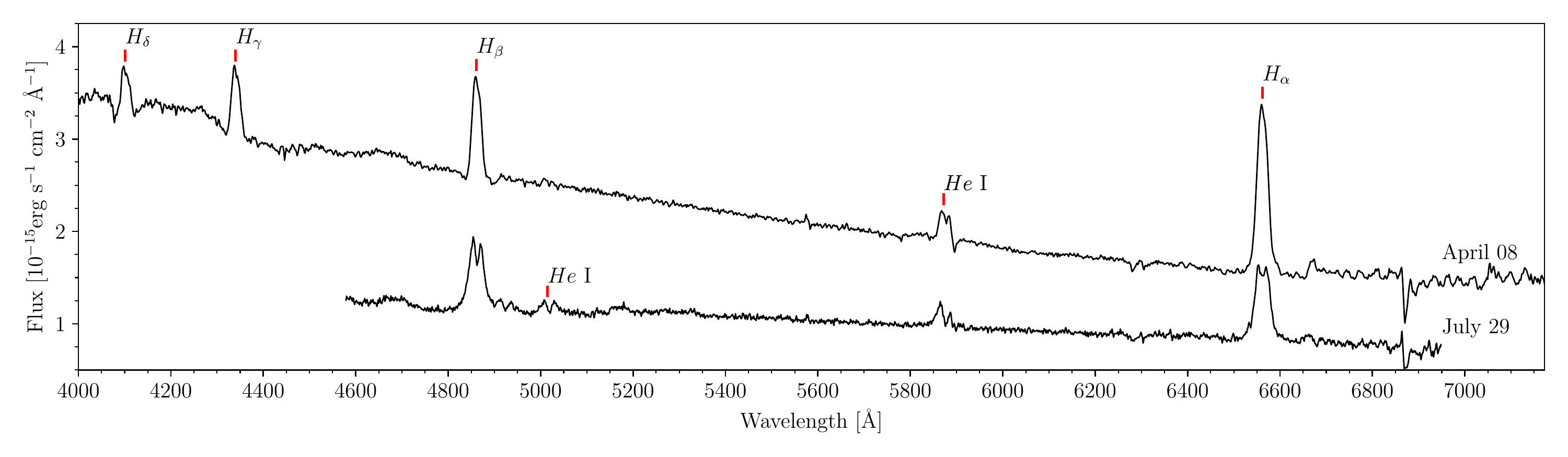}
        \caption{Low-resolution optical spectrum of CzeV404~Her averaged over one orbital period obtained during two different activity phase: during the outburst (2019, April 8) and  quiescence (2019, July 29). 
        The Balmer emissions   and the He~I 5884\AA\ and He~I 5016\AA\ lines are  marked. }
        \label{Fig:02}
\end{figure*}

\begin{figure*}[t]
        \centering
        \includegraphics[width=1.0\textwidth]{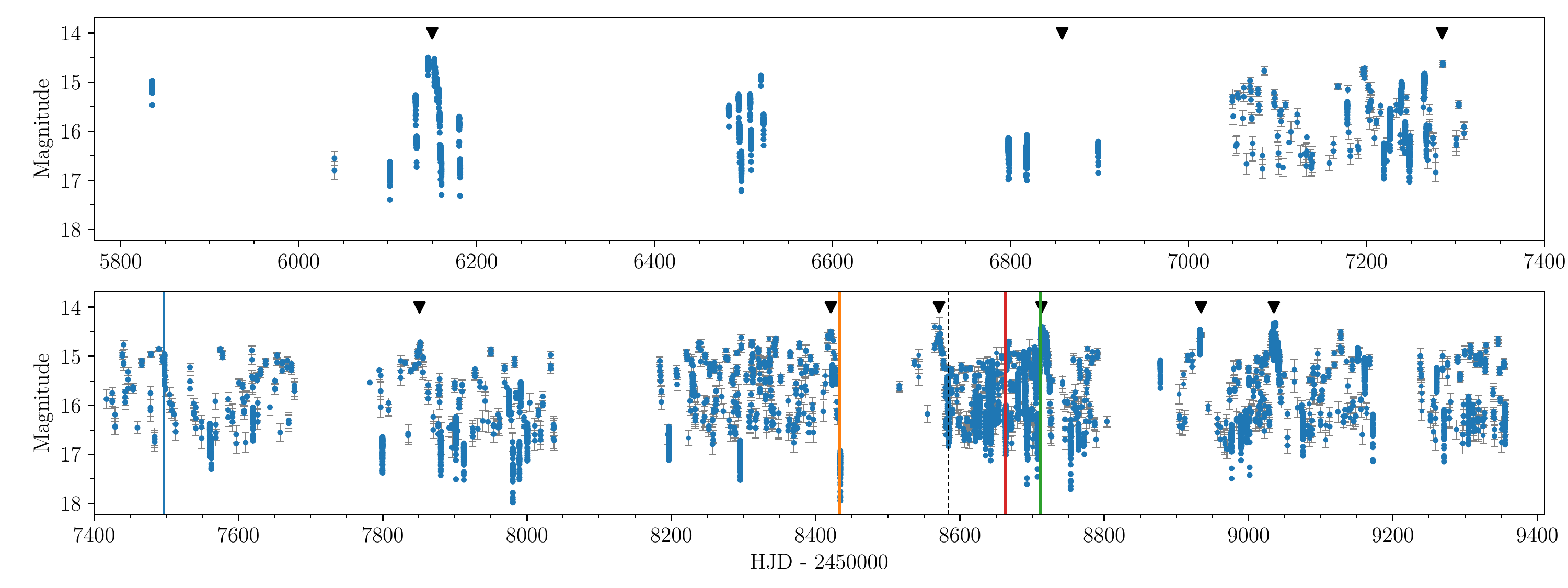}
        \caption{Observations of CzeV404~Her obtained in the ASAS-SN project, at the BSO, and at Ond\v{r}ejov Observatory. The V-band data from
        the ASAS-SN project were shifted by $+0.4$~mag, and the C-band  data of Ond\v{r}ejov  and  BSO  were shifted by $+0.2$~mag. Black triangles
        mark superoutbursts detected by these observations. Vertical full lines mark  the observation epochs  presented 
        in Fig.~\ref{Fig:06}. The same colour is used for the same observation time. Vertical dashed lines mark the time of spectroscopic observations listed in Table \ref{Table_02}.}
        \label{Fig:03}
\end{figure*}

\begin{figure} 
        \centering
        \includegraphics[width=0.48\textwidth]{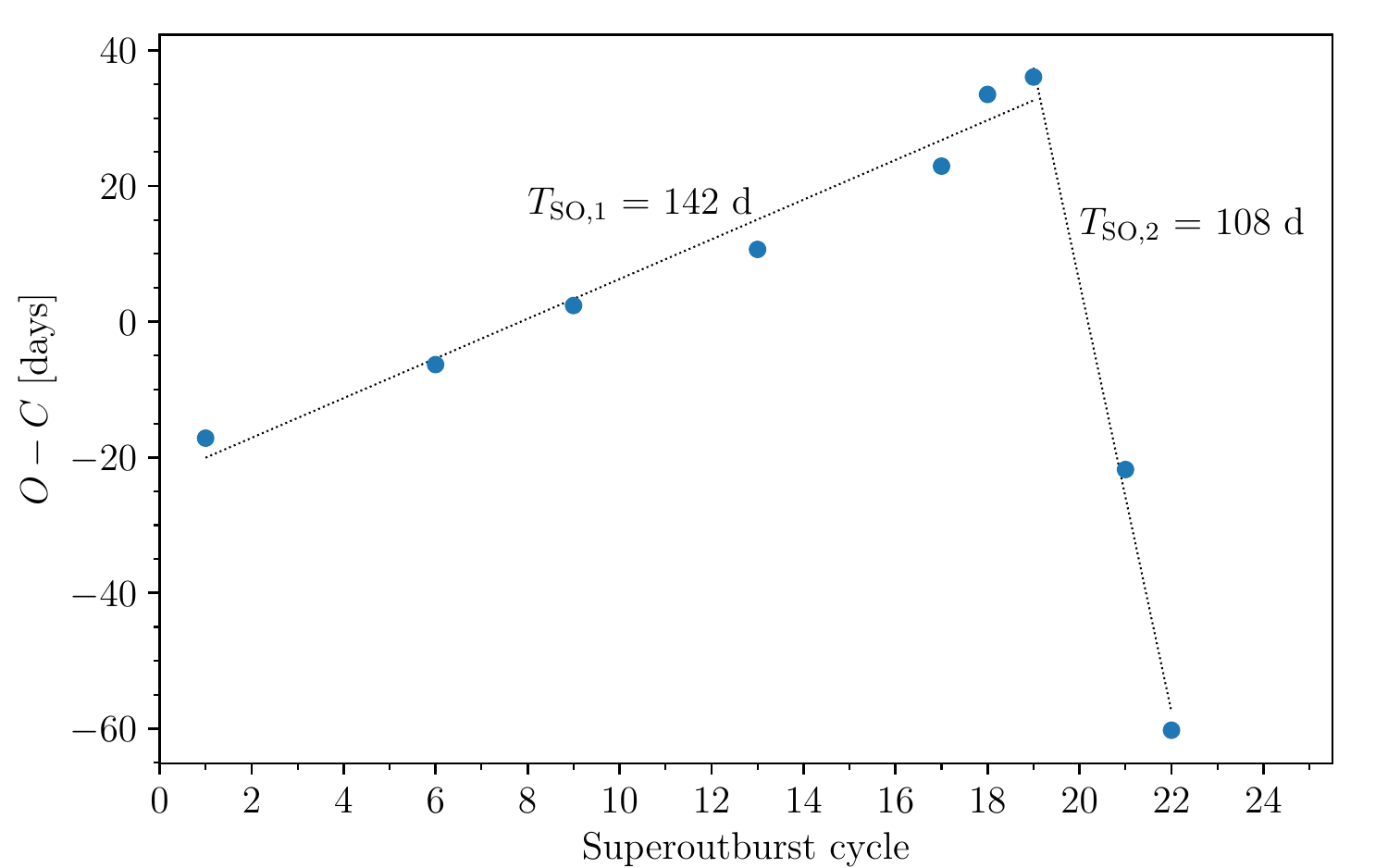}
        \caption{Observed minus calculated ($O-C$) diagram for superoutburst occurrences. The expected moments of superoutbursts were computed accepting a recurrent time of $T_{\mathrm{SO}} = 140$ days. Linear fits for intervals of 1 -- 19 and 19 -- 22 cycles are shown. The calculated average times between superoutbursts are marked for each interval.
        }
        \label{Fig:04.5}
\end{figure}

\begin{figure} 
        \centering
        \includegraphics[width=0.48\textwidth]{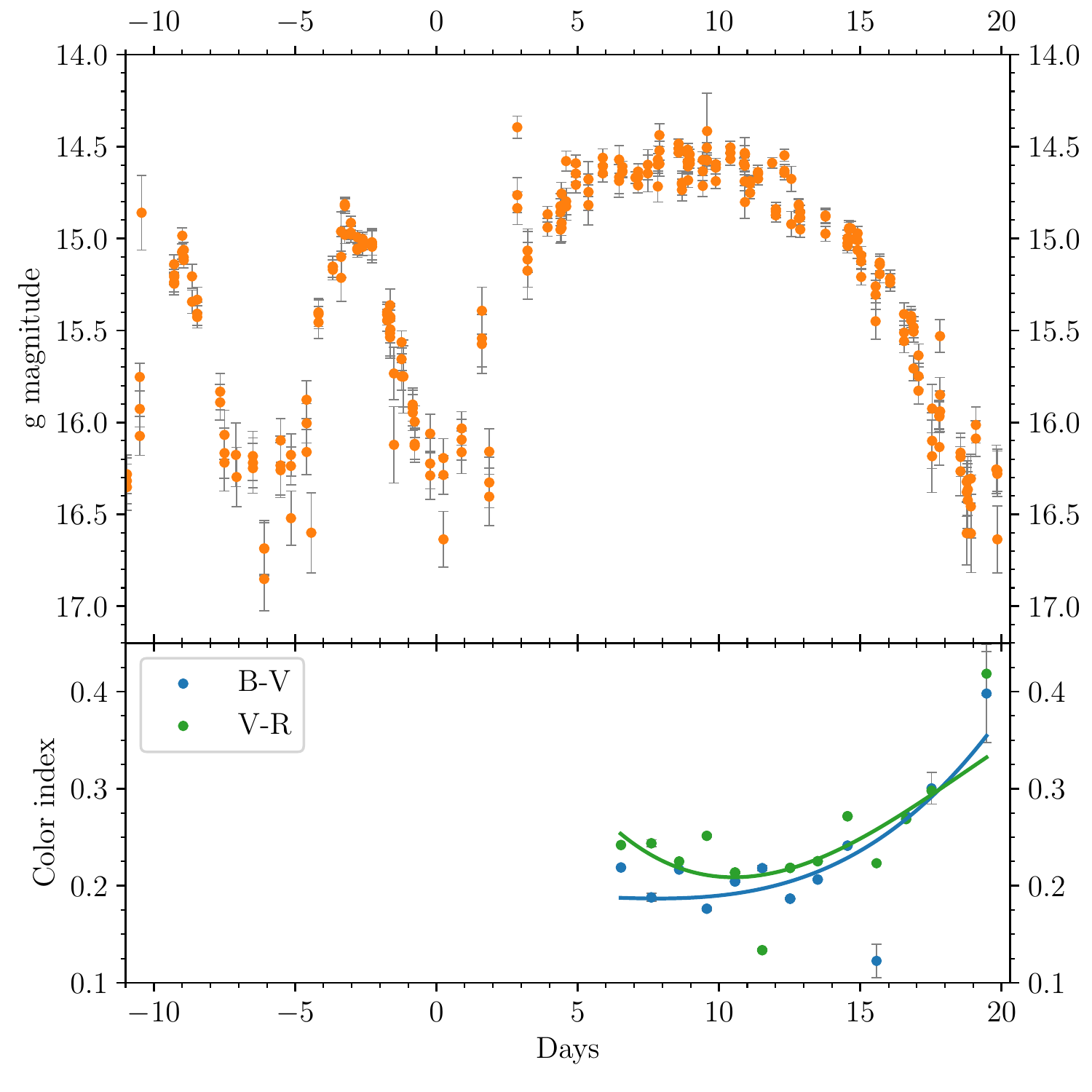}
        \caption{Zoom of the average  g-band light curve centred on a superoutburst  together with two preceding normal outbursts (top; see the description in the text).
        Evolution of colour indices $B-V$ (blue dots) and $V-R$ (green dots)  during the superoutburst 
        of the cycle 22 (bottom).}
        \label{Fig:04}
\end{figure}

\begin{figure*}
        \centering
        \includegraphics[width=0.98\textwidth]{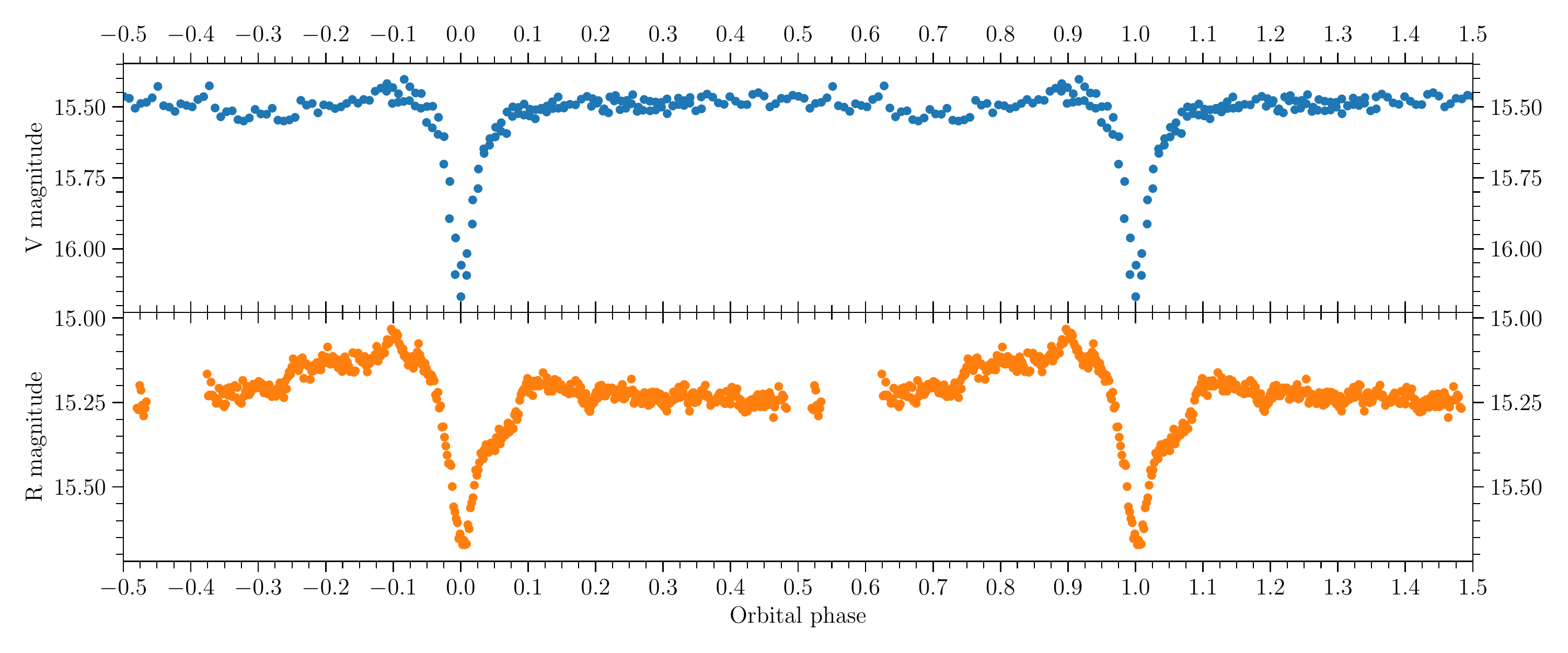}
        \caption{Light curves of CzeV404~Her folded on the orbital period. 
        The top panel shows the V-band light curve obtained at the SPM observatory 
        on 2019, April 9, and the bottom panel corresponds to the R-band  
        light curve obtained at the La Silla Observatory on 2019, March 15 (V=15.6). }
        \label{Fig:05}
\end{figure*}

The subject of this paper is the study of CzeV404~Her 
(also 2MASS~J18300176+1233462,  UCAC4~513-078584).
\cite{2014IBVS.6097....1C} reported the discovery of the object and determined its orbital period as $\sim$2.35~hours.
Based on the detection of the superoutburst and several outbursts, they classified the source as an SU~UMa-type CV.
The system resides inside the period gap and is currently known as the eclipsing  SU~UMa-type object with the longest orbital period.
 \cite{2014AcA....64..337B} 
 confirmed the SU~UMa nature of CzeV404~Her by detecting a superoutburst with an amplitude of $2\fm2$ that lasted  $\sim$17~days.  The superhump period $P_\mathrm{sh}$ = 0.10472(2)\footnote{Herein, the numbers in brackets are 1$\sigma$ uncertainties referring to the last significant digits quoted.} day was found  with a highly decreasing rate of 
$\dot{P_{\mathrm{sh}}}$ = -2.43(8)$\times10^{-4}$~s~s$^{-1}$.  
Based on 17 eclipses, they derived the orbital period of $P_{\mathrm{orb}}$ = 0.0980203(6) day.  The period excess was $\epsilon=0.068(2)$.  \cite{2014AcA....64..337B}
estimated  the mass ratio in the system to be $q\approx 0.32$\footnote{We note that the improved  $\epsilon\sim q$ relation \citep[see equation (8)]{2005PASP..117.1204P} gives $q=0.26$.  Equation (5) from \citet{2013PASJ...65..115K} gives a lower value of $q=0.21$.} based on the $\epsilon\sim q$ relation from \citet{1998PASP..110.1132P}.

\citet{2017MNRAS.465.4968H}  refined  the ephemeris of  CzeV404~Her:
\begin{center}
Pri. Min. = BMJD 24 56871$\fd$91730(4) + 0$\fd$09802125(2) $\cdot E$.
\end{center}

\noindent
The {\sl Gaia} parallax of the source  is $\pi=$2.94(5)~mas 
\citep{2018A&A...616A...9L}, which gives a distance of $d=340(6)$~pc.\footnote{
\cite{2018A&A...616A...9L} showed that when the relative error of the parallax is sufficiently low, inverting the parallax is a valid approach to obtain the accurate distance to the object. With a relative error $< 2 \%$ in the case 
of CzeV404~Her, inverting the parallax is a valid approach to determine the distance.} The interstellar absorption in this direction for the object distance is  $E(g-r)=0.17(2)$ \citep{2015ApJ...810...25G}. 

Here we report the results of new time-resolved photometric and spectroscopic observations of CzeV404~Her.  
The paper is structured as follows. In Sections~\ref{sec:phot} and \ref{sec:spec} we describe our
observations,  the photometric data available from the ASAS-SN
photometry database, and the data reduction. In Sections~\ref{sec:longphot} and
\ref{sec:orbphot} we analyse the outburst and superoutburst activity and
the short-term light curves of CzeV404~Her. The eclipse light-curve modelling and fundamental system parameters are presented in Section~\ref{sec:LCfit}. Section~\ref{Dopmaps} reports the accretion flow
structure based on the Doppler mapping method.  The results are discussed and our conclusions are given in Sections~\ref{sec:disc} and \ref{sec:conc}, respectively.

\section{Photometry}
\label{sec:phot}

The CCD photometry of CzeV404~Her was obtained at  different observatories: 
at Ond\v{r}ejov Observatory  (Czech Republic), 
at La Silla Observatory (Chile),  
at the Observatorio Astronomico Nacional San Pedro M\'{a}rtir (M\'{e}xico, SPM hereafter), 
at the Observatorio del Roque de los Muchachos (Spain, ORM hereafter), and at BSObservatory\footnote{\url{http://www.tcmt.org}} 
(Czech Republic, BSO hereafter), a private observatory in the Czech Republic operated by P. Caga\v{s}. The log of photometric observations is presented in Table \ref{Table_01}.
The field star 2MASS~J18295743+1233421 ($V = 15.465(2)$ mag, see Fig.~\ref{Fig:01}) was used as a standard. Its magnitude was calculated from the Pan-STARRS1 data \citep{2016arXiv161205560C}
using the transformation equations reported by \cite{2018BlgAJ..28....3K}.

Standard data reduction was carried out using the {\it apphot}/{\sc IRAF} tools for SPM data, the  {\sc Aphot32}~\footnote{developed at the Ond\v{r}ejov
Observatory by M.~Velen and P.~Pravec} and {\sc C-MUNIPACK}~\footnote{ \url{http://c-munipack.sourceforge.net}, created by D.~Motl} software in the case of data obtained at La Silla and Ond\v{r}ejov observatories, {\sc C-MUNIPACK} for ORM data, and 
{\sc SIPS}~\footnote{\url{https://www.gxccd.com/cat?id=146&lang=409},  created by P.~Caga\v{s} } for BSO data.

\begin{table}[t]
\caption{Log of the photometric monitoring of CzeV404~Her.}
\label{Table_01}
\centering
\begin{tabular}{c c c c c c}
\hline\hline\noalign{\smallskip}
Observatory  &  Photometric bands & Monitoring period \\
\hline\noalign{\smallskip}
Ond\v{r}ejov  & C, V, R    &  2015 --  2021  \\
SPM           & V          &  2019 \\
La Silla      & R          &  2019 \\
BSO           & C          &  2012 -- 2020  \\
ORM           & B, V, R    &  2020      \\
\hline 
\end{tabular} 
\end{table}

\subsection{Ond\v{r}ejov Observatory photometry}

Since June 2015, the CzeV404~Her eclipses have been regularly monitored at Ond\v{r}ejov Observatory in the Czech Republic. 
A total of 44 eclipse light curves has been collected using the Mayer 0.65 m ($f/3.6$) reflecting telescope with the CCD camera G2-3200 in clear light (called "C" filter) with an exposure time of 30~seconds. In addition, the series of $\sim$10$\times 60$ second 
frames in the V and R bands were carried out in 2019 -- 2020 in order to study long-term brightness variations of the system. 

\subsection{SPM photometry}

Time-resolved photometry of CzeV404~Her was obtained using the direct CCD image mode of the  0.84 m telescope
OAN-SPM Observatory in M\'{e}xico. The object was observed on 2019, April 9 -- 11, and on July 28 -- 29. The time of an 
individual exposure was 60 seconds and the runs lasted about 2 -- 3.5~hours.  The images were bias-corrected and flat-fielded before aperture photometry was carried out.
The photometric data were calibrated using the field standard star. The errors, ranging from 0.01 to 0.05 mag, were estimated from the
magnitude dispersion of comparison stars with similar brightness. One whole orbital period light curve and three additional light curves of the eclipse were collected. 
\begin{table*}[htb!]
        \caption{Log of spectroscopic observations. }
        \label{Table_02}
      \begin{center}
          
        \begin{tabular}{l c c c c c c c c}
        \hline\hline\noalign{\smallskip}
          Date  & HJD$^*$   & Range  &Dispersion & Exp.   & Number   &  Duration & V band  \\ 
           (2019).  & +2450000    &  [\AA] &  [\AA/pixel ] & time [min] & of spectra  &    [hour] & [mag] \\
         \hline\noalign{\smallskip}
           Apr 08 (N1)   & 8582.872  & $3700- 7200$ & $1.7$ &$10$ & $17$ &  $3.1$ & $15.5$ \\
           Apr 09 (N2)   & 8583.922  & $3700- 7200$ & $1.7$ &$10$ & $13$ &  $2.2$  & $16.0$\\ 
           Apr 10 (N3) & 8584.888  & $6000 - 7200$ & $0.6$ &$14$ & 13 &    $3.1$  & $16.5$\\ 
           \multirow{2}{*}{Jul 28~~~(N4)} & \multirow{2}{*}{ 8692.687 } &  \multirow{2}{*}{$6000 - 7200$} & \multirow{2}{*}{$0.6$} & $10$  & 8 &  $2.1$  & \multirow{2}{*}{$16.7$}\\   
             & & & & 15 & 10 &  3.4\\ 
          Jul 29~~~(N5) & 8693.753   &  $4600 - 6800$ & $1.2$ &$10$ & $22$ &  $4.0$ & $16.5$ \\
            \hline 
        \end{tabular} 
        \end{center}
\end{table*}

\subsection{La Silla photometry}
The  additional single-night light curve of CzeV404~Her was obtained  in the R band with a total observation 
time of~2 hours and an exposure time of 7~s on 2019, March 15. 
The Danish 1.54 m telescope located at La Silla Observatory 
in Chile was used.

\subsection{BSO photometry}
 We also used data from the long-term monitoring in the clear light collected during 
 2014 -- 2019 using the 25 cm and  30 cm  reflecting telescopes of  BSObservatory in the Czech Republic.

\subsection{ORM photometry}
The BVR-bands observations of the superoutburst (cycle~22, Table~\ref{sup_tab})  and the  out-of-eclipse brightness monitoring of the object were obtained during 2020, July, using the FRAM~CTA-N 25 cm telescope located at the Observatorio del Roque de los Muchachos, La Palma, Canary Islands, Spain \citep{2019arXiv190908085E}. 

\subsection{ASAS-SN photometry database}
The ASAS-SN photometry database\footnote{\url{https://asas-sn.osu.edu/}} \citep{2014ApJ...788...48S, 2017PASP..129j4502K,2019MNRAS.485..961J}
contains the V-band (2012, 2015 -- 2018) and g-band (2018 -- 2021)  photometric data of CzeV404~Her.
We extracted the data using the ASAS-SN photometry database online tool. A circular aperture of 16$^{\prime\prime}$ radius was
used for the aperture photometry. Several background stars fell into the aperture and contaminated the object flux.  We verified
that none of these stars showed any variability and  added a constant value to the source flux measurements.  The ASAS-SN data typically cover a short time interval during one night (up to three points per night).  They are  only suitable for studying long-term variability and monitoring normal and superoutbursts.

\begin{table}[htb!]
    \caption{List of superoutbursts}
        \label{sup_tab}
        \centering
    \begin{threeparttable}
        
        \begin{tabular}{llc}
            \hline\hline\noalign{\smallskip}
             Cycle & HJD\tnote{a} & Source \\ 
            \hline\noalign{\smallskip}
           1    &   2456150 & \cite{2014IBVS.6097....1C}  \\
           6    &   2456858 & \cite{2014AcA....64..337B}  \\
           9    &   2457285 & ASAS-SN V    \\
           13   &   2457851 & ASAS-SN V  \\
           17   &   2458421 & ASAS-SN g  \\
           18   &   2458571 & ASAS-SN g  \\
           19   &   2458713 & Ond\v{r}ejov, ASAS-SN g \\
           21   &   2458934 &  ASAS-SN g \\ 
     \multirow{2}{*}{22}       &  \multirow{2}{*}{2459035} & Ond\v{r}ejov, ASAS-SN g   \\    
                                &           &   ORM  BVR                \\
        \hline 
        \end{tabular} 
        \begin{tablenotes}
        \item[a]  HJD of the maximum brightness
        \end{tablenotes}
        \end{threeparttable}
\end{table}

\begin{figure*}[!h]
        \centering
        \includegraphics[width=0.48\textwidth]{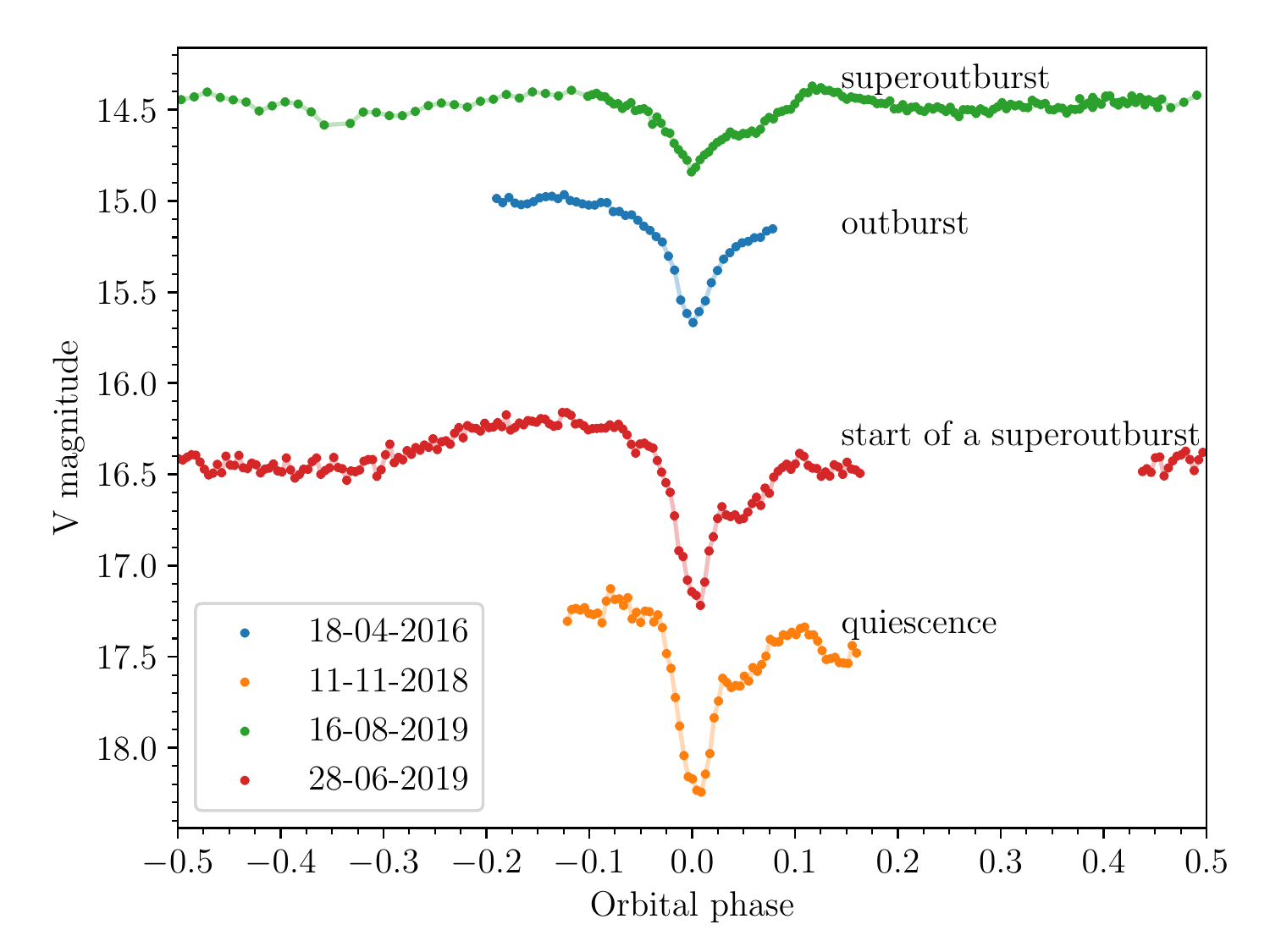}
        \includegraphics[width=0.48\textwidth]{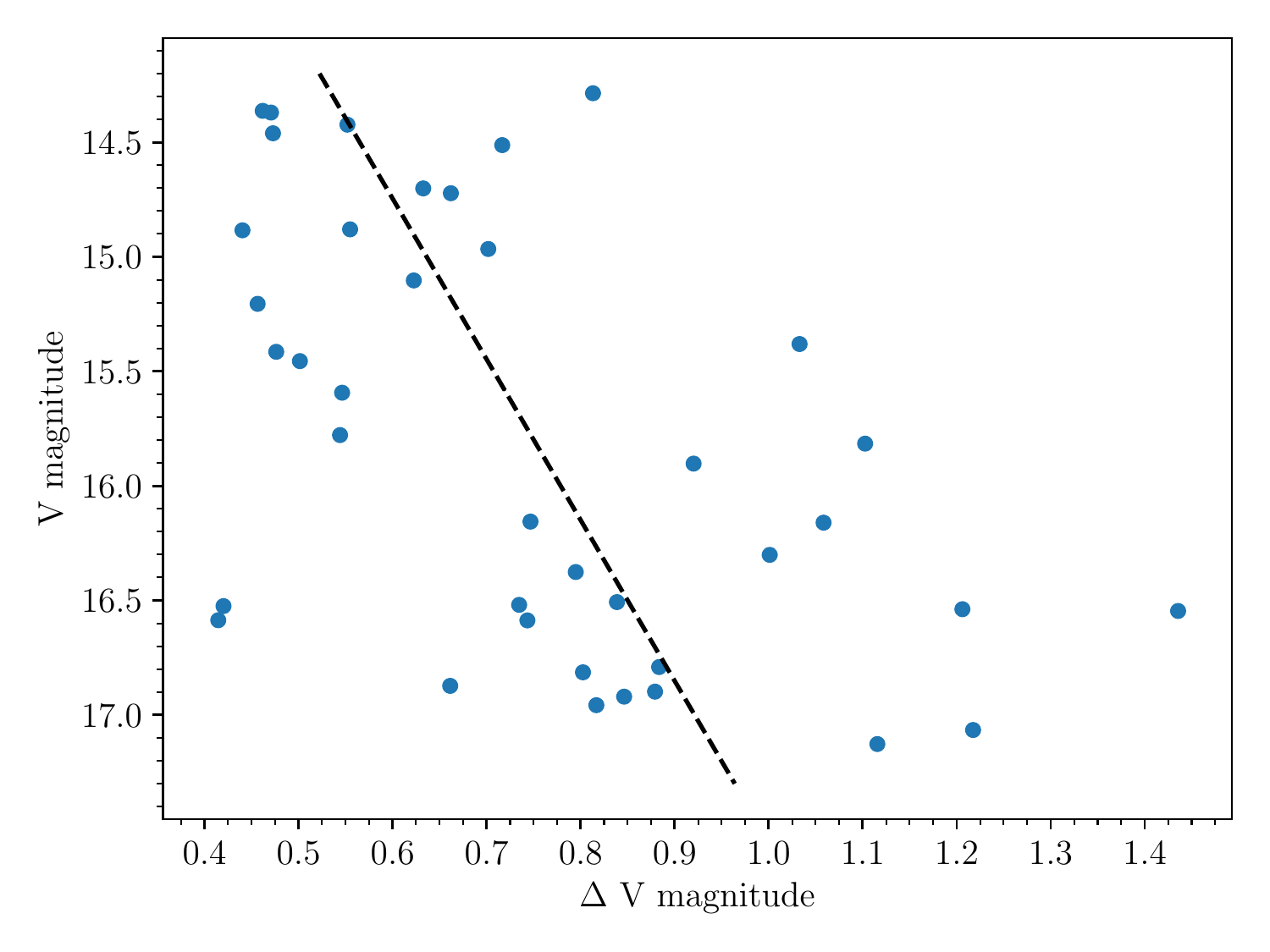}
        \caption{Eclipsing light curves obtained at different state of  CzeV404~Her (left). 
        Relation of the eclipse depth vs. the out-of-eclipse brightness of the system (right). The dashed line shows a linear fit to the data.
        }
         \label{Fig:06}
\end{figure*}

\begin{figure}[!htb] 
        \centering
        \includegraphics[width=0.48\textwidth]{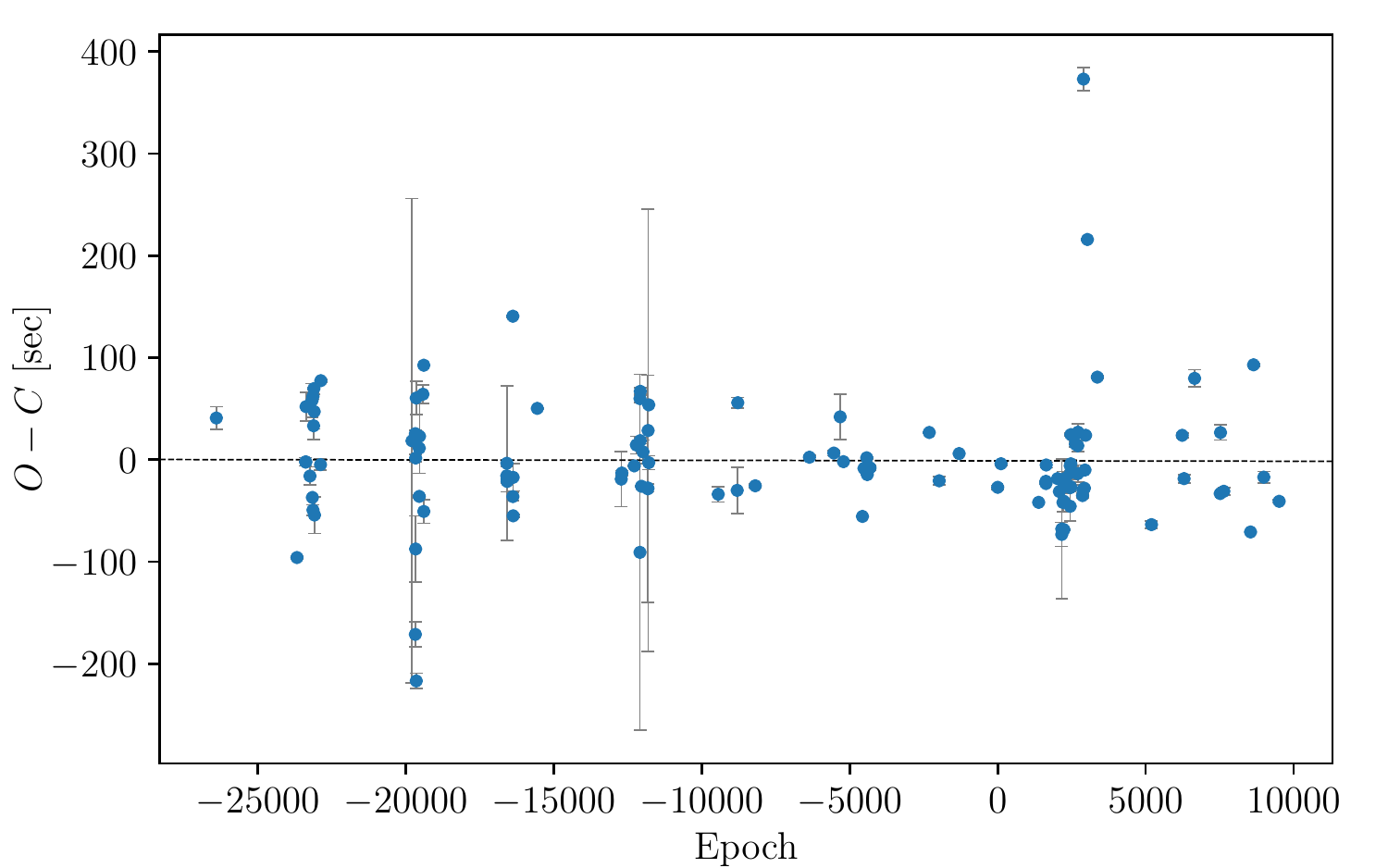}
        \caption{
        Observed minus calculated ($O-C$) diagram for the times of the primary eclipse
        of CzeV404~Her. The errors bars are indicated. No significant variations  in the $O - C$ values are noted.
        }
        \label{Fig:07.5}
\end{figure}

\begin{figure*}
\setlength{\unitlength}{1mm}
\resizebox{11cm}{!}{
\begin{picture}(100,125)(0,0)
\put (0,0)  {  \includegraphics[width=15.0cm, bb = 50 0 680 510,clip=]{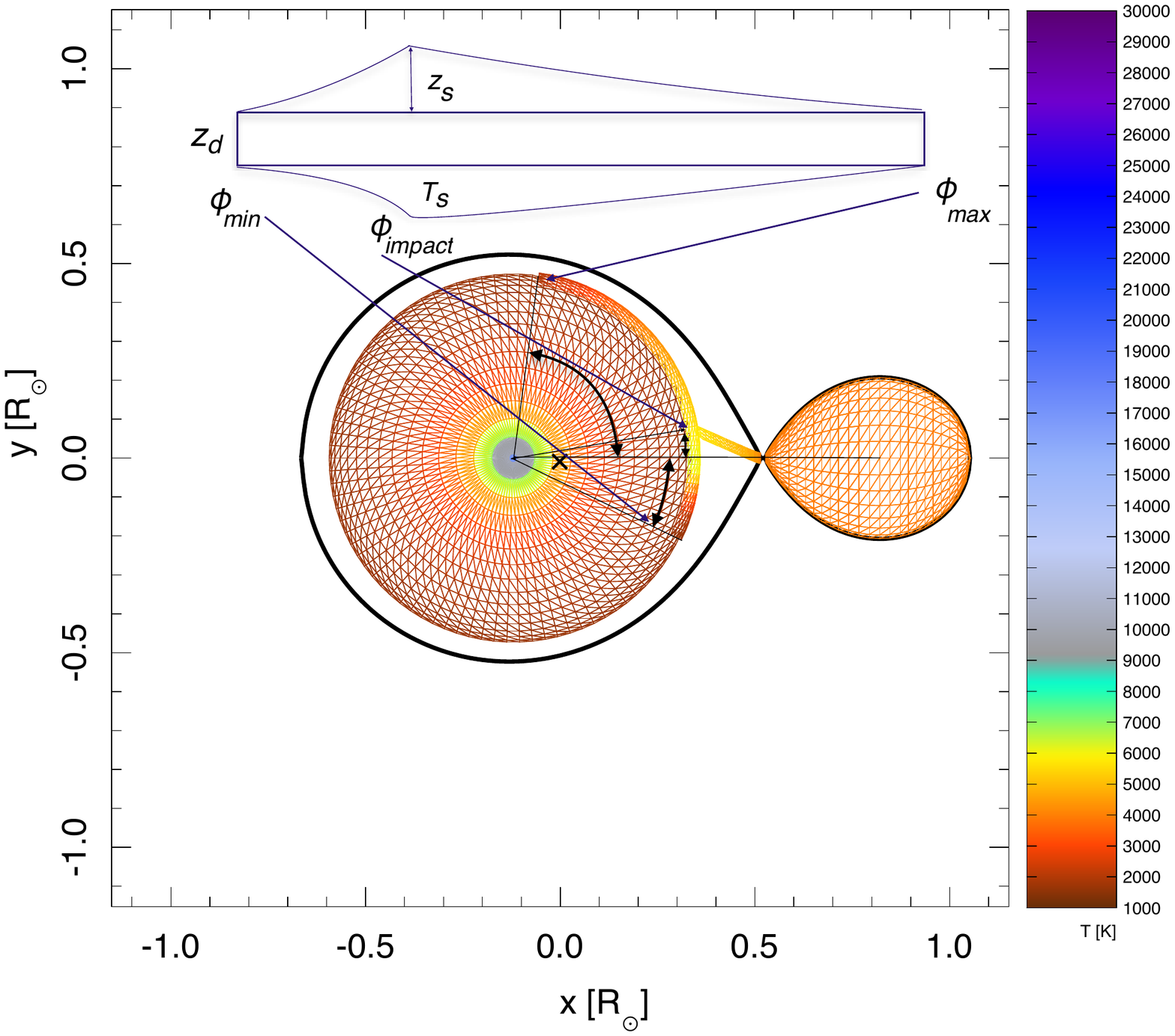} }
\end{picture}}
\caption{Model of CzeV404~Her including the definition of the hot-spot parameters. See the full description of the parameters in the text (Section~\ref{sec:LCfit}). $z_d \equiv h_{\mathrm{d,out}}$; $\phi$ is the angle from the line connecting the centres of the primary and secondary. $T_s$ and $z_s$ correspond to maximum values of the hot-spot size in z-direction and its effective temperature.
}
\label{Fig:07}
\end{figure*}

\section{Spectroscopy}
\label{sec:spec}

Long-slit spectroscopic observations were obtained using the Boller \& Chivens spectrograph attached to the 2.12 m 
telescope of the SPM during two observing runs. The first run was on 2019,  April 8 -- 10 and the second run on 2019, July 28 -- 29. 
The weather conditions during the first run 
were photometric, but the second run was partly affected by a small  variation in the sky transparency. 
The spectral data were reduced in the standard way using the {\it apextract} and {\it onedspec } IRAF tools.
The log of spectroscopic observations is given in  Table~\ref{Table_02}.
The period-averaged spectra of the source obtained in 2019, 
April and July, are shown in Fig.~\ref{Fig:02}. 
The spectra clearly confirm the CV nature of the object.

\section{Long-term behaviour of CzeV404~Her}
\label{sec:longphot}

Fig.~\ref{Fig:03} shows long-term behaviour of  CzeV404~Her on a timescale of about 3000 days. The object varies strongly between 14\fm5--16\fm5  in the V and g bands. 
Normal outbursts and superoutbursts can be clearly distinguished in the light curve. The normal outbursts of CzeV404~Her have an amplitude of $A_{\mathrm{NO}}\sim 2\fm0$ in the V band, and they occur with a recurrence period of $T_\mathrm{NO}\sim7$~days. A normal outburst lasts about $5$ days. 
The object brightness increases from quiescence to the maximum with a rate of $\sim1\fm0$~day$^{-1}$ , and the duration of the maximum source brightness is about $1$~day.  After the maximum, the object falls to quiescence with a rate that is twice slower ($\sim0\fm5$ day$^{-1}$) than the increasing phase.

\begin{table*}[htb!]
\centering
\caption{List of observed mid-eclipse times.}
\label{table_OC}
\begin{tabular}{l l l l c}
\hline\hline\noalign{\smallskip} 
 HJD -- 2400000 &  BJD -- 2400000       &       r.m.s.      &  Epoch & Observatory\\
\hline\noalign{\smallskip}
55835.352271    &       55835.353038    &       0.000127        & -26401        & BSO \\
56102.458581    &       56102.459355    &       0.000013    & -23676    & BSO \\
56131.375919    &       56131.376705    &       0.000002        & -23381        & BSO \\
...             &   ...             &   ...         &   ... \\
\hline 
\end{tabular} 
\tablefoot{The full table is available in the electronic form.}
\end{table*}

Various superoutbursts were also observed by the ASAS-SN project in addition  to those reported by \cite{2014IBVS.6097....1C} and \cite{2014AcA....64..337B} (see Table~\ref{sup_tab}). 
The amplitude of a superoutburst in the V band is  $A_{\mathrm{SO}}\sim 2\fm4$, which exceeds the amplitude of a normal outburst by about  $\sim 0\fm4$. A superoutburst lasts about 20~days. Typically, the brightness increases with a rate of 0\fm5 day$^{-1}$ during  $\sim 5$~days, then the accretion disk spends $\sim 8$ days in a plateau phase, and after this, the brightness declines for $\sim 7$~days  to quiescence. All superoutbursts recorded so far are listed in Table~\ref{sup_tab}.
The mean recurrent time based on all superoutbursts is $T_\mathrm{SO} = 140$ days. This recurrent time was used to compute the cycle numbers and to construct the $O-C$ diagram for the superoutburst occurrence presented in Fig.~\ref{Fig:04.5}. The $O-C$ diagram shows that the recurrent time changed after the cycle 19.
We estimate that superoutbursts occurred with a recurrent time of  $T_\mathrm{SO} = 142$~days between epochs 
of HJD 2456150 -- 2458713.  
The occurrence times of cycles 21 and 22 imply a recurrent time $T_\mathrm{SO} \sim 108$~days, but no superoutbursts were observed after cycle 22 to support this value.
Two events were observed at HJD 2459127\footnote{observed
in the ASAS-SN project} and  HJD 2459255\footnote{based on an alert available at
\url{http://ooruri.kusastro.kyoto-u.ac.jp/mailarchive/vsnet-alert/25366}}. The epochs of these events correspond to a recurrent time of  $\sim$108 days after cycle 22. Nevertheless, although the amplitudes of brightness increase were comparable to a superoutburst, their durations were the same as of a normal outburst. 

In Fig.~\ref{Fig:04} we plot the fragment of the light curve centred on a superoutburst together with two preceding normal outbursts. The light curve was formed by a combination of data of all superoutbursts observed by the ASAS-SN project in g band. The data were shifted arbitrarily so that the shapes of the superoutbursts are outlined.  
 We note that the amplitudes ($A_{\mathrm{NO}}$, $A_{\mathrm{SO}}$) and the periods ($T_\mathrm{NO}$, $T_\mathrm{SO}$) of outbursts and superoutbursts 
of CzeV404~Her follow the corresponding linear relations $A_{\mathrm{NO,SO}}$ versus $T_{\mathrm{NO,SO}}$  derived by \cite{2016MNRAS.460.2526O}
for SU~UMa-type stars.

\section{Orbital period light curves of CzeV404~Her} 
\label{sec:orbphot}

In Fig.~\ref{Fig:05} we display typical light curves of the object folded on the orbital period. The  magnitudes out of eclipses were $V\sim15\fm5$  (2019, April 9) and
$R\sim15\fm2$  (2019, March 15).  They correspond to intermediate states of the system  (Fig.~\ref{Fig:04}). 
There is  flickering in the light curve. The increase in the brightness of the object before  eclipses is related to the hot spot at the edge of the accretion disk. The  eclipse depth was about $\Delta V\sim0\fm7$  and $\Delta R \sim 0\fm5$. We note that the shape and  depth of the eclipses depend on  the system brightness (Fig.~\ref{Fig:06}).

To search for orbital period variations, we measured moments of the minima of eclipses using the {\sc lcurve} software \citep{2010MNRAS.402.1824C}.  The results  are listed 
in Table \ref{table_OC} and were used to update the system ephemeris,
\begin{center}
Pri. Min. = BJD 24 58423$\fd$21151(8) + 0$\fd$098021247(6) $\cdot E$.
\end{center}
No significant variations  in the orbital period are noted (Fig.~\ref{Fig:07.5}). 
The standard deviation of the moments of eclipse minima is  $\sigma$ = 52~s.

\begin{table}
\centering
\caption{System parameters from in the light-curve modelling of CzeV404~Her in quiescence.
}
\label{tab:BestPar}
\begin{tabular}{llll}
\hline\hline\noalign{\smallskip}
{\bf Fixed parameters:}       &            &          \\  
\hline\noalign{\smallskip}
\multicolumn{2}{l}{$P_{\mathrm{orb}}$}  &  \multicolumn{2}{c} {8469.04 s}   \\ 
\multicolumn{2}{l}{$E(B-V)$ }           &  \multicolumn{2}{c} {0.16(3)}    \\
\multicolumn{2}{l}{Distance}            &  \multicolumn{2}{c} {340(6) pc}  \\    
\hline\noalign{\smallskip}
\multicolumn{4}{l}{{\bf Variable and their best values:}} \\  \hline\noalign{\smallskip}

\multicolumn{2}{l}{ $i$ }               & \multicolumn{2}{c}{ 78\fdg8(4) } \\
\multicolumn{2}{l}{$M_{\mathrm{WD}} $ } &\multicolumn{2}{c}{1.00(2) M$_{\sun}$ }  \\
\multicolumn{2}{l}{${T}_{\mathrm{WD}}$ }     & \multicolumn{2}{c}{17000(5000)  K}\\ \noalign{\smallskip}\hline\noalign{\smallskip}
\multicolumn{2}{l}{$\dot{M}$} & \multicolumn{2}{c}{2.03(3)$\times10^{-11}$M$_{\sun}$ year$^{-1}$}\\
\multicolumn{2}{l}{$T_{2}$ }   & \multicolumn{2}{c}{$ 4100(50)$  K}\\

\hline\noalign{\smallskip}
\multicolumn{4}{l}{{\bf Parameters$^*$ of the disk:}} \\ 
\hline\noalign{\smallskip}
\multicolumn{2}{l}{ $R_{\mathrm{d, in}}$  }  & \multicolumn{2}{c}{ 0.008 R$_{\sun}$ } \\
\multicolumn{2}{l}{ $R_{\mathrm{d, out}}$ } & \multicolumn{2}{c}{ 0.47(2) R$_{\sun}$ } \\
\multicolumn{2}{l}{ $h_{\mathrm{d, out}}$ } & \multicolumn{2}{c}{ 0.024(5)  R$_{\sun}$ } \\  

\noalign{\smallskip}\hline\noalign{\smallskip}
\multicolumn{2}{l}{{\bf The hot spot/line:}}& \multicolumn{1}{c}{Q} & \multicolumn{1}{c}{S} \\ \hline\noalign{\smallskip}
\multicolumn{2}{l}{length spot ($\varphi_{\mathrm{min}}+ \varphi_{\mathrm{max}}$) }  &\multicolumn{1}{c} {111\fdg0(5)} & \multicolumn{1}{c}{28\fdg0(5)} \\
\multicolumn{2}{l}{width spot (\%) }  &\multicolumn{1}{c} {5.0(2.5)} & \multicolumn{1}{c} {0.0(1)} \\
\multicolumn{2}{l}{Temp. excess spot ($T_{\mathrm{s, max}}/T_{\mathrm{d,out}}$)  }  &\multicolumn{1}{c} {2.9} &\multicolumn{1}{c} {8.5} \\
\multicolumn{2}{l}{Shift  spot ($\varphi_{\mathrm{min}}$)}  &\multicolumn{1}{c} {-21\fdg9$(7)$} &\multicolumn{1}{c} {-10\fdg0(4)}  \\
\multicolumn{2}{l}{Shift $T_{\mathrm{s, max}}$ ($\varphi(T_{\mathrm{s, max}})$)}  &\multicolumn{1}{c} {-12\fdg5(2)} &\multicolumn{1}{c} {-2\fdg3(1.0)}  \\

\hline\noalign{\smallskip}
\multicolumn{4}{l}{{\bf Calculated:}} \\ 
\hline\noalign{\smallskip}
\multicolumn{2}{l}{$a  $}  &\multicolumn{2}{c}{0.94  R$_{\sun}$}  \\
\multicolumn{2}{l}{$q  $}  & \multicolumn{2}{c}{ 0.159}  \\
\multicolumn{2}{l}{$M_{2}  $}  & \multicolumn{2}{c}{ 0.158 M$_{\sun} $} \\
\multicolumn{2}{l}{$R_{2}$ }     & \multicolumn{2}{c}{0.221  R$_{\sun}$}\\
\multicolumn{2}{l}{$R_{\mathrm{WD}}$ }     & \multicolumn{2}{c}{0.008  R$_{\sun}$}\\
\hline

\end{tabular}
\tablefoot{Numbers in brackets for the variables are uncertainties defined as the interval in which the $\chi^2$-function increases twice compared with the minimum value. The distance and $E(B-V)$ are given with 1$\sigma$ errors.
The letters "Q" and "S" mark quiescence and the superoutburst parameters of the hot spot or line, respectively. }
\end{table}

\section{Eclipse light-curve modelling}
\label{sec:LCfit}

We used the collected photometric observations to analyse the eclipse light curves  using a tool developed by \citet{2013A&A...549A..77Z} to 
define the system parameters. Briefly, the model  includes a primary white dwarf, a secondary red dwarf star, a stream of accretion
matter, an accretion disk with thickness $z_\mathrm{d}(r) = z_\mathrm{d}(r_{\mathrm{out}})(r/r_{\mathrm{out}})^\gamma$ , and  the standard  effective temperature gradient 
\begin{equation}
\begin{split}
T_\mathrm{d}(r)  \sim T_* \times r^{-3/4}, {\mathrm{where}~~}  
T_*  = \left(\frac{3GM_{\mathrm{WD}}\dot{M}}{8\pi \sigma R_{\mathrm{WD}}^3}\right)^{1/4},
\end{split}
\end{equation}
\citep[see ][]{1995CAS....28.....W} between the inner and outer edges ($r_{\mathrm{out}} \approx 0.60 a /(1+q)$) of the disk, 
and a hot spot or line. The white dwarf is a sphere, defined by the mass-radius relation in \citet[2.83b]{1995Ap&SS.226..187W}. The secondary 
is assumed to fill its Roche lobe, and the Roche-lobe shape is directly calculated using Equation 2.2 \cite[]{1995Ap&SS.226..187W} for
equipotential $\Phi$(L$_1$). The surface of each component of the system is divided into a series of triangles. We assume that each 
triangle emits as a black body with the corresponding effective temperature. The limb-darkening \citep{2012A&A...546A..14C} and 
the illumination of the secondary by the primary are also included. The intensity of each element is convolved with the corresponding 
filter bandpass and converted into the flux taking into account the element surface, the orientation to the line of view of an observer,
the distance to the system, and the interstellar absorption. The light curves of individual components and the binary system as a 
whole were obtained by integrating the emission from all  elements lying in view of an observer. 

Free parameters of the fit are the mass of the primary ($M_{\mathrm{WD}}$), the mass ratio ($q\equiv M_2/M_{\mathrm{WD}}$), 
the mass transfer rate ($\dot{M}$),  the system inclination ($i$), the outer radius of the accretion disk
($R_{\mathrm{d, out}}$),
the disk thickness at the outer radius ($h_{\mathrm{d, out}}$),  
the effective temperature of the secondary ($T_2$), and the parameters of the hot spot or line. 
Because the hot spot is a dominating source in the continuum (also in the Balmer emission lines, see Section~\ref{Dopmaps} below) 
around the eclipse, we used a complex model for the spot, as presented in Fig.~\ref{Fig:07}.  The impact of the stream with 
the outer edge  of the disk forms a shock-heated trailing arc along the rim. The vertical size $z_{\mathrm s}$ and the temperature 
$T_{\mathrm s}$ of the arc-like hot spot are  not uniform; it is wider and  hotter at the impact point ($\varphi_{\mathrm{impact}}$), 
quickly increases before the impact point,  and slowly declines after one  in the temperature and the vertical extension, 
according to an arbitrarily selected manner described by the following equations:
\begin{equation}
\begin{split}
 T_{\mathrm{s},i}(\varphi)& =T_{\mathrm{d}}(1.0+\gamma_{T,i}f(\varphi)^{\alpha_{T,i}}) \\
 z_{\mathrm{s},i}(\varphi) & =z_{\mathrm{d}}(1.0+\gamma_{z,i}f(\varphi)^{\alpha_{z,i}}),
 \end{split}
 \end{equation}
 where $\varphi$ is  the angle between the line connecting the stars in conjunction and the direction of the spot viewed from 
 the centre of mass; $f(\varphi) \equiv a\times\varphi+b$ is a linear function with constants $a$ and $b$ defined by boundary conditions 
 of $f(\varphi_{\mathrm{min}})  = 0$, $f(\varphi_{\mathrm{impact}})  = 1$ and  $f(\varphi_{\mathrm{max}})  = 0$, and  $\gamma_T,\gamma_z$ are  free parameters. The index $i = 1,2$ corresponds to parts of the hot spot before and after the impact point.

\begin{figure*}[h]\setlength{\unitlength}{1mm}
\resizebox{11cm}{!}{
\begin{picture}(100,135)(0,0)
\put (0,70)  {\includegraphics[width=0.48\textwidth]{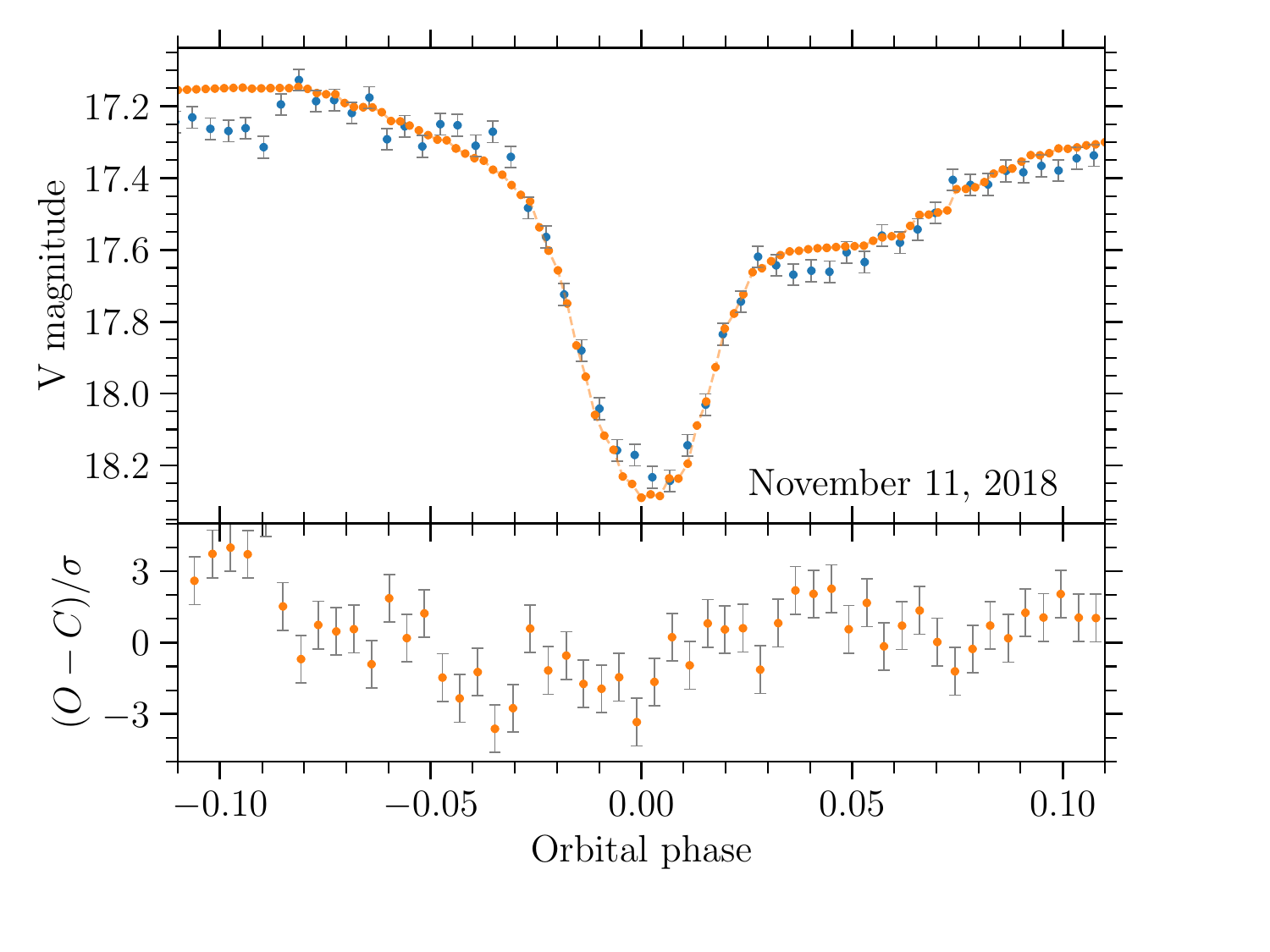}} 
\put(80,70)  {\includegraphics[width=0.48\textwidth]{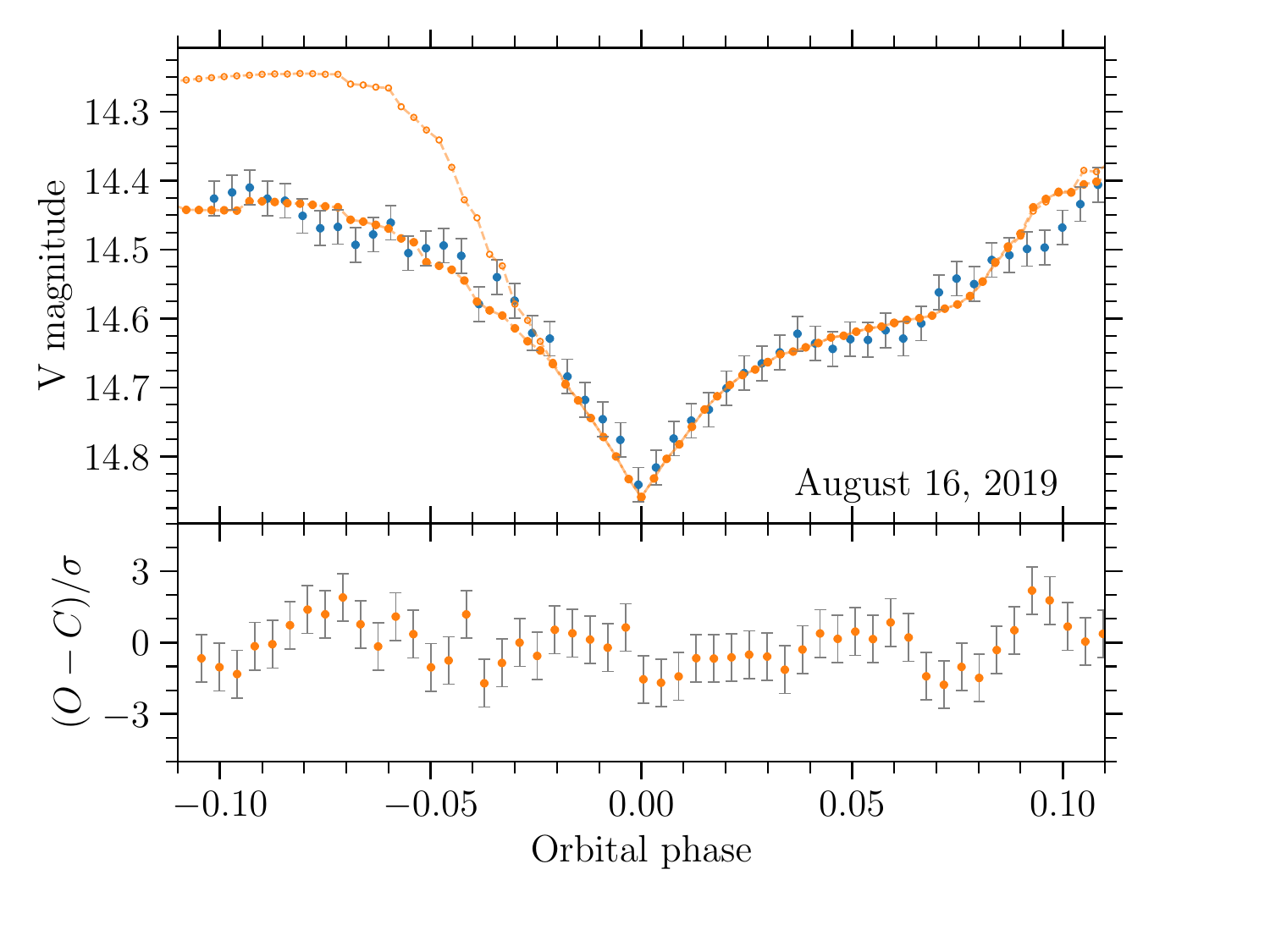}} 
\put (0,0)  {\includegraphics[width=0.44\textwidth,bb = 0 0 1370 1250, clip=]{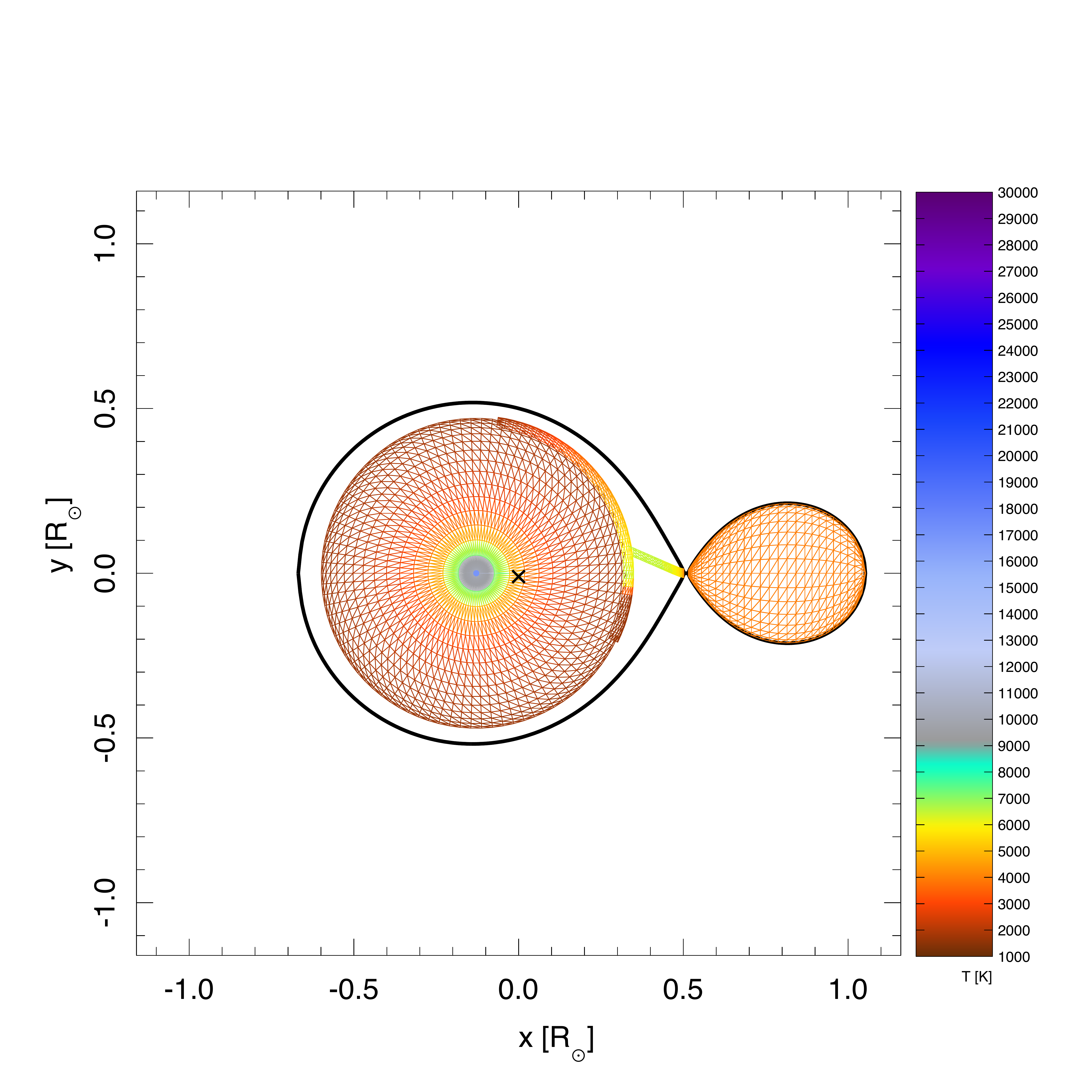}} 
\put(81,0)  {\includegraphics[width=0.48\textwidth,bb = 0 0 1500 1250, clip=]{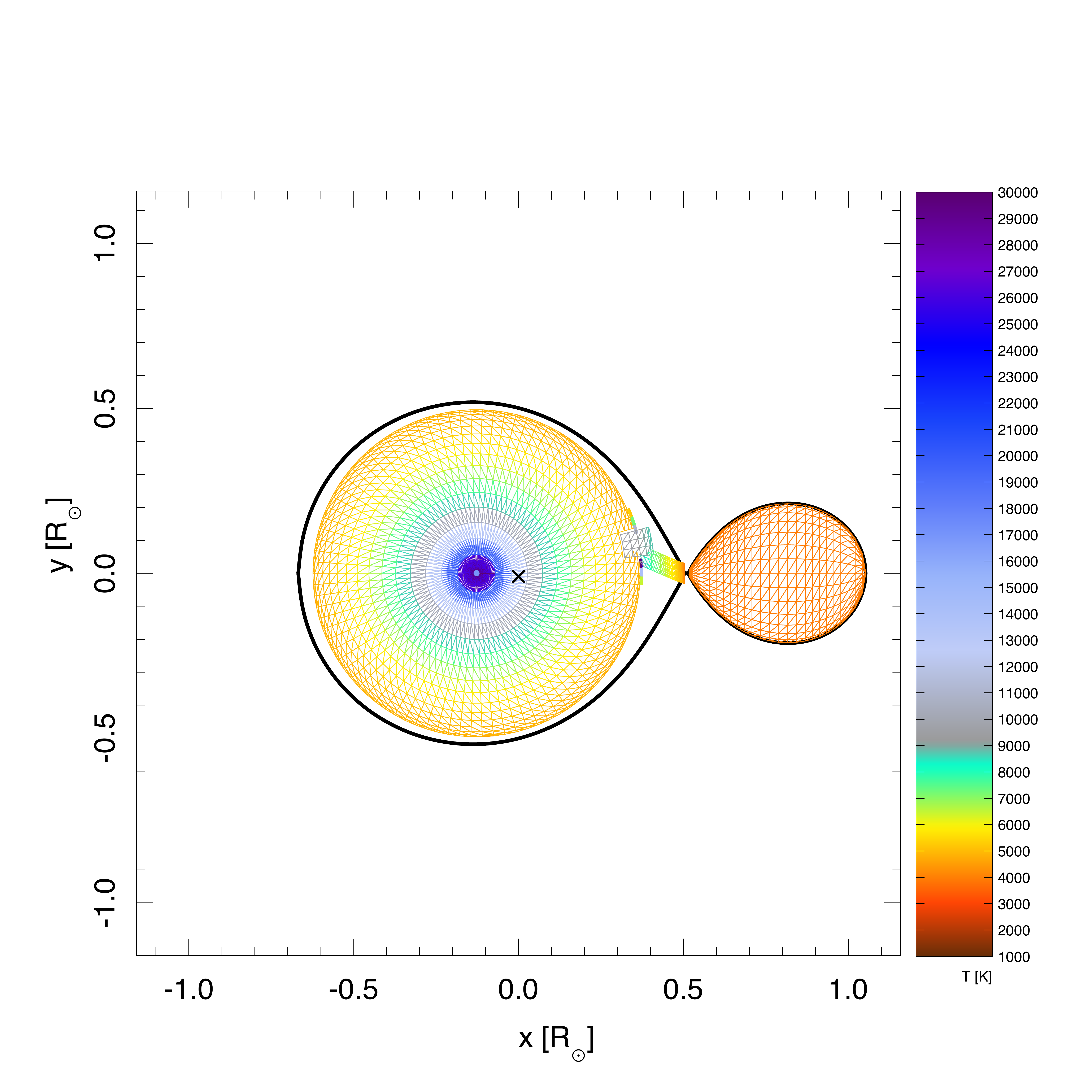}} 
\end{picture}}
 \caption{Results  of the light curve fit of CzeV404~Her eclipses in quiescence (top left) and in superoutburst (top right). 
 The blue points show the observational data presented in Fig.~\ref{Fig:06}. The orange points mark the modelling light curves corresponding to the best fits. The $O-C$ residuals for the fits are given. Models of CzeV404~Her in quiescence (bottom left) and in outburst (bottom right). The colours mark the effective temperature of the system elements (right bar).   }
\label{Fig:08}
\end{figure*}

 The eclipses of  CzeV404~Her presented in Fig.~\ref{Fig:06},~left, were used to fit the system parameters beginning with 
 the light curve obtained in  quiescence  (2018, November 11). Only the phase region [-0.09, 0.09] was used for 
 the fitting where only weak effects of the flickering related to an inner variability of the disk or/and the spot are expected.
 We searched for the minimum of the $\chi^2$ function with the gradient-descent method. It is defined as 
\begin{equation}
 \chi^2 = \sum_k^{N_k}\left(\frac{mag_{\mathrm{obs},k}-mag_{\mathrm{calc},k}}{\sigma~mag_{\mathrm{obs},k}}\right)^2 ,    
\end{equation}
\noindent
where $N_k$ is the number of observed points in each fitted light curve. The values of the best-fit model are presented in
Table~\ref{tab:BestPar}. The error  of each fitted parameter  was calculated with the Gaussian 
approximation of the $\chi^2$ function when other parameters were  fixed  at the best values. 
The result of the fit and the observed minus calculated $(O-C)$ residuals are  shown in Fig.~\ref{Fig:08},~top left panel, and the corresponding geometry and temperature distribution of the system components are given in  Fig.~\ref{Fig:08},~bottom left panel.   

\begin{figure*}
\setlength{\unitlength}{1mm}
\resizebox{11cm}{!}{
\begin{picture}(100,145)(0,0)
\put (0,0)  {  \includegraphics[width=16.5cm, bb = 20 180 550 680,clip=]{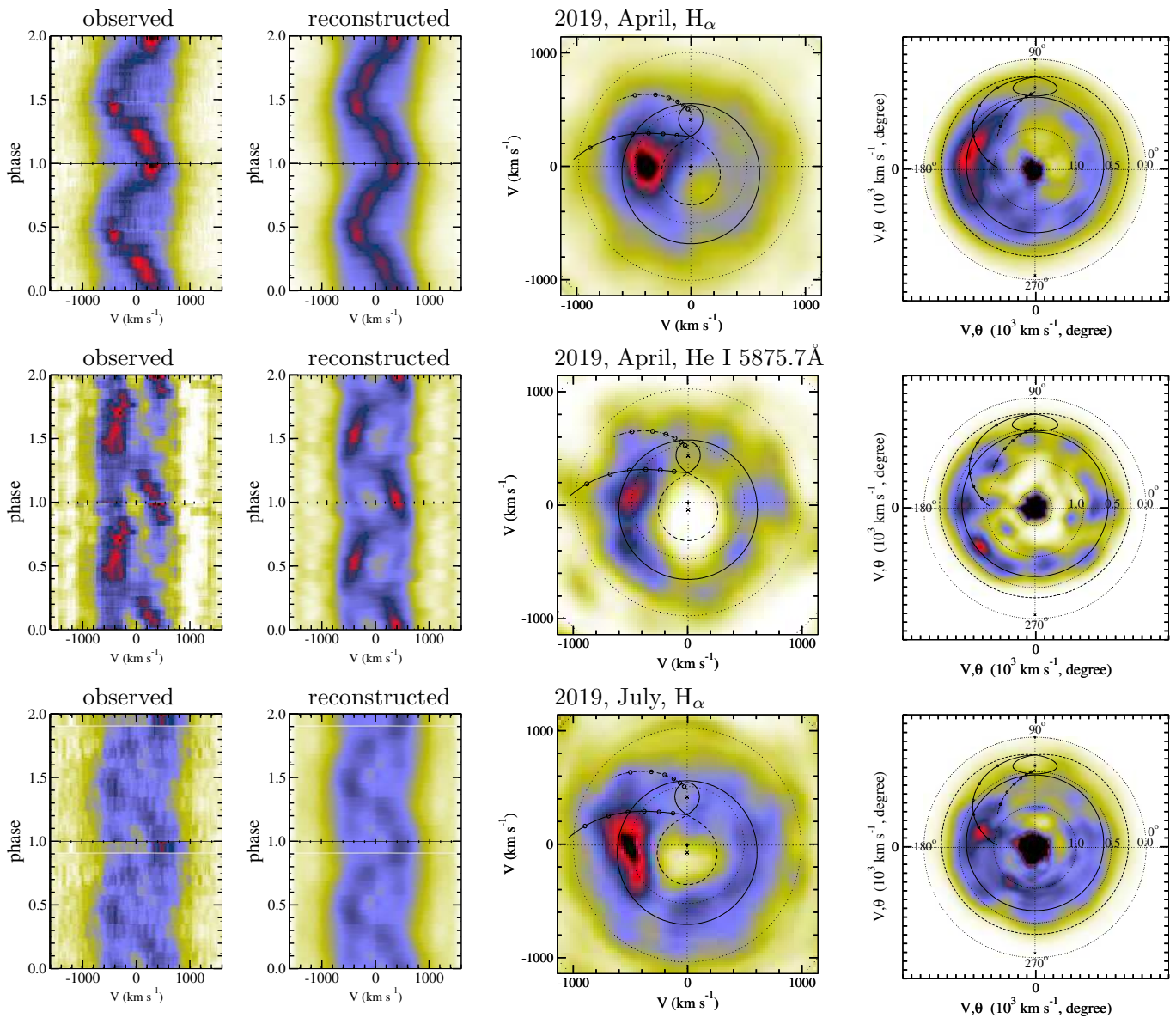} } 
\end{picture}}
\caption{ Trailed observed and reconstructed spectra of the H$_\alpha$ and He I emission lines  (left two panels) and the standard 
and inside-out projection Doppler maps (right two panels) of Czev404~Her obtained in  April 2019 (top, middle panels) and 
July 2019 (bottom panels). The circle (solid line) in the Doppler maps marks the tidal limitation radius of the accretion disk. 
The central spot in the inside-out projection is an artefact.}
\label{Fig:09}
\end{figure*}

Using the same values for the most of the parameters, we calculated (Fig.~\ref{Fig:08},~top right) the model eclipse light curve (filled orange circles)  of CzeV404~Her for the superoutburst on 2019, August 16 (blue circles).  The observed eclipse light curve was  reproduced by an increase in mass transfer rate up to
$\dot{M} = 1.46\times10^{-9}M_\sun$~year$^{-1}$  together with changes in the spot parameters and configuration 
(see  the Table~\ref{tab:BestPar}),  and  the disk slope $\gamma$ = 0.078 in the $z\sim r^{\gamma}$ relation. 
The main difference between the model of the quiescence and the superoutburst eclipse light curves is a substantial change in
the mass transfer rate (of about two orders). The disk in a superoutburst is significantly hotter and practically flat, with a constant size along the z-axis independent of its radial size.
This means that the disk thickness becomes wider in the inner part during a superoutburst.
The impact region has a higher temperature, but the total contrast of the spot within the disk becomes low, and it appears to have a notably smaller geometrical size in the model.  The stream between the secondary and the disk is optically thick and affects the eclipse light curve because it absorbs part of the radiation from the hot spot. An additional element is needed in the model (see
Fig.~\ref{Fig:08},~bottom right) to absorb part of the radiation of the hot spot.
We show non-absorbing (open circles) and self-absorbing  (filled orange circles) models of light curves of the system for comparison in Fig.~\ref{Fig:08},~top right.  
 We interpret this additional element as matter that bounces off the disk during impact of the stream.
Having a temperature slightly lower than that of the edge of the disk, it covers part of the hottest spot region for an observer at eclipse ingress.  This results in the lower brightness of the spot 
compared to what is expected when the observed flux from the spot is extrapolated after the eclipse to the orbital phases preceding it.
The outer radius of the accretion disk does not change much between quiescence and outbursts. 
Other very important results of eclipse light-curve modelling are a relatively high mass of the white 
dwarf, $M_{\mathrm{WD}} \approx 1.0$ M$_{\sun}$\ , and that the secondary with a temperature 
$T_2=4100 $~K is hotter than expected for a main-sequence star with a mass of 0.16~M$_{\sun}$. The last two factors are probably responsible for the fact that this object is observed inside the period gap. Clearly, the parameters of the spot are only effective parameters of the model. They show how the observed behaviour of the object can be described within the model framework.  
The exact description of the impact area and its observational manifestations are more complex. Describing them requires a complete non-conservative magnetohydrodynamic modelling of the disk and the disk-stream impact region.

\begin{figure}
\setlength{\unitlength}{1mm}
\resizebox{11cm}{!}{
\begin{picture}(110,80)(0,0)
\put (0,0)  { \includegraphics[width=8.5cm, bb = 150 150 1450 1430,clip=]{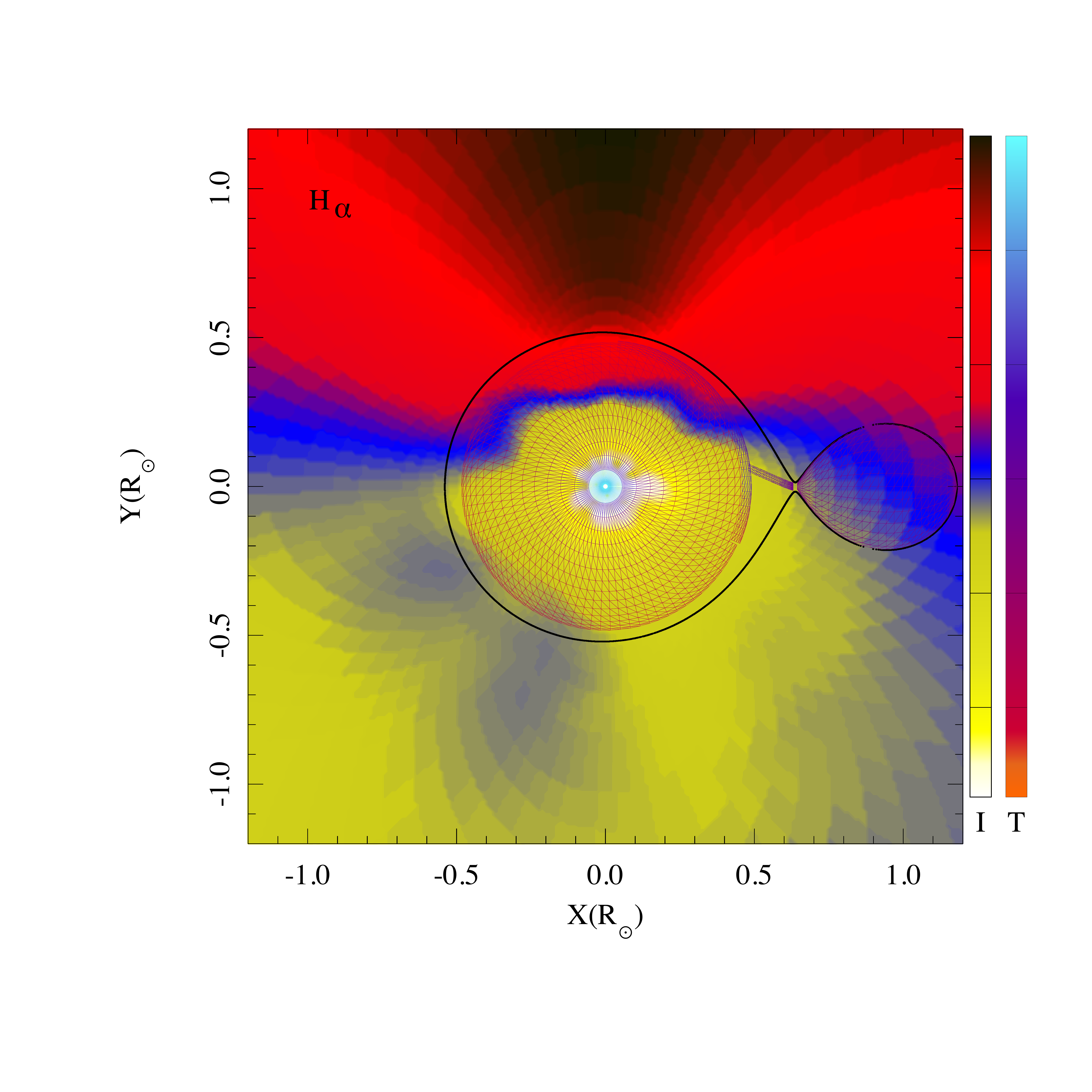} } 
\end{picture}}
\caption{Brightness distribution "I" transformed from the Doppler map of the H$_\alpha$ emission line to the XY plane of the system together with the geometrical view "T" of the system in effective temperatures (1000--10~000~K), as follows from the light-curve modelling.}
\label{Fig:10}
\end{figure}

\section{Doppler tomography}
\label{Dopmaps}

We used the Doppler tomography technique \citep{1988MNRAS.235..269M}  to probe the accretion flow structure in CzeV404~Her. This technique is a well-known method to map the emission of gas moving in the  $(V_x, V_y)$ velocity plane. It was applied to map the  H$\alpha$ and He~I emission lines.
We generated Doppler maps from our time-resolved spectra of the object using the maximum entropy method  \citet{1998astro.ph..6141S}
 as implemented by \citet{2016yCat..35950047K}. We combined the data from nights N1--N3 (2019, April) and N4+N5 (2019, July; see Table~\ref{Table_02}). The first set was obtained during a decline of a normal outburst, while the second one was observed close to the quiescence level. The resulting trailed spectra and Doppler maps are shown in Fig.~\ref{Fig:09}.
The orbital phase zero was well defined from the simultaneous photometry obtained in the same nights as the spectroscopy. 
The possible displacement between orbital phase zero when the secondary stays strictly between the primary and an observer and a minimum of an eclipse does not exceed $\sim$0.005 orbital phase, as follows from our light-curve modelling estimate.

The trailed spectra of the H$_\alpha$ line in 2019 April clearly show a low-velocity S-wave with an amplitude of about 
 200 km~s$^{-1}$. The S-wave varies in intensity with orbital phase. It is brighter at 0.0 -- 0.5 orbital phases and diminishes significantly at phases of 0.5 -- 1.0. There is a dip in intensity at the phase about 0.25.
The system parameters from Section~\ref{sec:LCfit}  were used to draw the primary and secondary Roche lobes, 
the ballistic and Keplerian stream trajectories, and the centre of mass of the system in the Doppler maps. 
The outermost radius of the disk is marked by the solid line circle at $V=630$~km~s$^{-1}$ in Fig.~\ref{Fig:09} and corresponds to the Keplerian velocity at the tidal limitation radius of the disk. 

The maps  look  unusual for an SU~UMa-type system\footnote{see, for examples, the Doppler maps of SU~UMa-type systems in \citet{2005A&A...431..269R, 2006A&A...451..613M, 2015MNRAS.447..149L, 2018AJ....155..232L}} presenting a bright spot at the low velocity of $V_x\approx -200$~km~s$^{-1}$, $V_y\approx 0$~km~s$^{-1}$ 
in the standard, and the $V\approx 600$~km~s$^{-1}$, $\theta \approx 170^\circ$ in the  inside-out projections. 
The disk-stream impact hot spot is in SU~UMa-type systems typically located inside the disk at or between the continuation of ballistic or Keplerian trajectories. However, in the case of Czev404~Her, the velocity of the spot does not correspond to the expected position. It is below the lowest Keplerian velocity in the accretion disk.

The disk is poorly expressed in the standard projection compared to the inside-out view (the spot in the centre of the maps is an artefact of the inside-out projection).   A similar picture is observed in He~I trailed spectra and Doppler maps where the spot and disk are less pronounced. The brightness variation of the S-wave in
He~I is also different from that in H$_\alpha$. The S-wave varies with the same phases, but has two dips in the intensity at orbital phases close to 0.25 and 0.8. In data from July 2019, when the object brightness was about one magnitude lower than in April 2019, we obtained a similar H$_\alpha$ map, but at lower intensity levels for its details. From this fact, we conclude that only the brightness of the accretion disk changed, but the accretion flow structure remained the same.

In Fig.~\ref{Fig:10} we plot the result of transforming the Doppler map calculated from April 2019 data into the system of spatial coordinates using the assumption of the Keplerian velocity distribution of the emitting particles. The colour bars marked by the letters "I" and "T" correspond to the intensity of the Doppler map similar to Fig.~\ref{Fig:09} and the effective temperature of the system components (from 3000~K to 10000~K), respectively.   The peak of the intensity (black region) is located out of the accretion disk, as also follows from the Doppler map. This fact can be interpreted to mean that the hot-spot region deviates strongly from Keplerian velocity.  Nevertheless, the part of the disk related to the tail of the impact region between the stream and the outer edge of the disk has a significantly higher intensity than the rest of the disk. This extended hot spot or line was proposed by different authors \citep{2013MNRAS.428.3559D, 2014AJ....147...68T} to explain the origin of SW~Sex-type systems.

\begin{table*}

\caption{Parameters of periodicity, outbursts, and superoutbursts of the selected period-gap CVs.  }
\label{T:OB_comp}
\begin{tabular}{l c c c c c c c c c}
\hline\hline\noalign{\smallskip}
 Parameter                              &  EF Peg       &       YZ Cnc          &       V344 Lyr     &       GZ Cnc  &       UV Gem  & V725 Aql      &   NY~Ser      &         CzeV404~Her        &            TU Men  \\ \hline\noalign{\smallskip}                                                                                                                                   
$P_{\rm orb}$ [d]               &       0.0837  &       0.0868          &       0.087904        &       0.0881  &       0.0895  &       0.0939  &       0.0978  &       0.0980                  &               0.1172  \\
Distance [pc]           &   255(26)     &       240(3)      &   1064(86)        &       181(11) &       667(79) &       746(285)&       773(31) &         340(6)                  &               278(5)  \\
$P_{\text{sh}}$ [d]     &       0.087   &       0.0903      &   0.0918          &       0.0928  &       -               &       0.0985  &       0.106   &       0.10472                 &               0.126   \\
$V_{\text{min}}$ [mag]  &       18.50   &       14.8            &       19.5            &       16              &       18.50   &       19.0    &       18.5    &       17.5                    &               18.50   \\
$V_{\text{NO}}$ [mag]   &       -               &       11.9            &       14.50           &       13.8    &       15.50   &       -               &       15.0    &       15.0                    &               12.50   \\
$V_{\text{SO}}$ [mag]   &       10.70   &       11.2            &       14.00           &       12.8    &       14.70   &       13.6    &       14.7    &       14.5                    &               11.60   \\ \hline
$T_{\mathrm{NO}}$ [d]           &       -               &       7                       &       10                      &       10              &       54              &       -               &       6               &       7                 &               37              \\
$T_{\mathrm{SO}}$ [d]       &   -               &       125                     &       260                     &       260             &       180             &       500             &       85              &       140                 &               150             \\
$D_{\text{NO}}$ [d]     &       -               &       4                       &       6                       &       6               &       -               &       -               &       4               &       5                            &           6               \\
$D_{\text{SO}}$ [d]     &       -               &       20                      &       17                      &       17              &       -               &       -               &       21              &       20                         &       -                       \\
$D_{\text{R}}$ [d]      &       -               &       4                       &       2                       &       2               &       -               &       1               &       3               &       5                         &       -                       \\
$D_{\text{P}}$ [d]      &       -               &       13                      &       8                       &       8               &       -               &       11              &       11              &       10                         &       -                       \\
$D_{\text{D}}$ [d]      &       -               &       3                       &       7                       &       7               &       -               &       -               &       7               &       5                                 &       -                       \\
\hline 
\end{tabular} 
\tablefoot{Data for all stars except for CzeV404~Her were published by
\citet[see the definition in  Fig.~1 therein and in the text]{2016MNRAS.460.2526O}, and distances are based on {\it Gaia} data \citep{2018yCat.1345....0G}. The superhump period of CzeV404~Her was adopted from \cite{2014AcA....64..337B}.
}
\end{table*}

\section{Discussion}
\label{sec:disc}

About 120 objects lie inside of the period gap (2.15 -- 3.18~h) and are proposed to be CVs following the last version 7.24 (2017, February) of the \citet{2003A&A...404..301R} catalogue. About 35\% of these CVs are polars and intermediate polars. Another 41\% objects are marked as DNe, 4 of them at the upper period-gap limit are classified as U~Gem-type systems, and the rest are SU~UMa-type objects. Five objects in the period range of 2.44 -- 3.05~h also show SW~Sex characteristics. Four objects in the list are eclipsing systems. One is the subject of this study, another is the recently discovered poorly studied DNe-type object 1RXS~J003828.7+250920 \citep{2016Ap.....59..304P}. The 1RXS~J003828.7+250920 shows outburst and superoutburst behaviour. The average light-curve profile of the object resembles the typical light curve of U~Gem and indicates multi-component emission by sources that appear to be the hot and cold parts of the accretion disk and a hot spot on the disk. Unfortunately, no spectroscopy  is available for this object.

Another two eclipsing  objects are nova-likes systems.
The first is V348~Pup, which was interpreted as an SW~Sex system \citep{2001MNRAS.328..903R, 2003AJ....126..964F} and is discussed below. The second is V~Per (Nova Persei 1887), a classical nova system \citep{1989ApJ...339L..75S}. The light curves of CzeV404~Her, V348~Pup, and V~Per   are very similar, and their spectra show single-peaked Balmer and He emission lines. V~Per is a faint ($V\approx 18$ mag) object with a larger accretion disk that shows a flat temperature distribution along its radius \citep{2006ApJ...644.1104S}.  No outburst activity is detected. 
Only limited spectroscopy has been presented so far  \citep{2007IBVS.5751....1H}, and a possible SW~Sex nature of the object has been indicated.  
Most of the other DNe-type objects in the period gap that have been proposed as SU~UMa systems are based on detected superoutbursts and/or superhumps, but they have not been studied in quiescence and only very poorly in outbursts.  Their system parameters are unknown, and long-term photometric or spectroscopic monitoring is missing.  Nevertheless, several period-gap DNe show outburst activity similar to CzeV404~Her.

The outburst characteristics of these objects are presented in Table~\ref{T:OB_comp}, 
where $P_{\rm  orb}$ and $P_{\rm sh}$ are orbital and superhump periods, respectively.
The next three rows give the average magnitudes of the systems in quiescence $V_{\rm min}$ and at
the maximum brightness of a normal, $V_{\rm NO}$, and  a superoutburst, $V_{\rm SO}$.
The marks $D_{\rm NO}$ and $D_{\rm SO}$ denote the duration of normal and superoutbursts, respectively.
The last is defined as $D_{\rm SO} = D_R + D_P + D_D$ as well, where $D_R$, $D_P$, and $D_D$ are the duration of the rise, the plateau phase, and the decline of an outburst.
The recurrent time of normal and superoutbursts are labelled  $T_\mathrm{NO}$ and $T_\mathrm{SO}$.  
All these values are given in units of days.   Table~\ref{T:OB_comp} shows that the outbursts and superoutbursts of CzeV404~Her have small amplitudes and a short recurrent time compared to other SU~UMa-type CVs placed in the period gap. 

 The spectra of SU~UMa systems in the period gap show some diversity. For example, YZ~Cnc,  GZ~Cnc, and EF~Peg  display one-peak relatively strong  Balmer and He~I emission lines accompanied by faint He~II $\lambda$ 4686\AA\ and Fe II $\lambda$ 5169\AA\ lines \citep{1988AJ.....95..178S,1994IBVS.4073....1H, 2002ApJ...575..419H, 2007PASP..119..494S} over a blue continuum. On the other hand, NY~Ser only shows low-intensity H$_\alpha$ emission at a blue continuum and probably faint high-order Balmer absorptions with indications of emission in its central part \citep{1998A&AS..128..277M}. However, no time-resolved spectroscopy and Doppler tomography is available for these objects. As a result, the structure of the accretion flows in these systems is unknown. 
Nevertheless, there is one well-studied  eclipsing system, V348~Pup, with an orbital period of 0.1018~days that is close to CzeV404~Her. The object was interpreted as an SW~Sex system in the period gap, as we mentioned above.
The spectrum of V348~Pup \citep{2001MNRAS.328..903R} shows single-peak Balmer and He~II emission lines. The He~I emission lines look like double-peak lines in some orbital phases. The intense He~II lines are evidence of a high-temperature region in the accretion disk. The trailed spectra of H$_\alpha$ emission and the Doppler map \citep[see their fig.~6]{2001MNRAS.328..903R} demonstrate a similar configuration as observed in CzeV404~Her (Fig.~\ref{Fig:09}). It is interesting that the radial velocity amplitudes of the emission lines in CzeV404~Her and V348~Pup are also very close to each other (200--300~km~s$^{-1}$).

The distance to CzeV404~Her is about 2.4 times smaller than the distance to V348~Pup (813(20)~pc). However, V348~Pup has $V\approx 15\fm0$ out of eclipse and does not show outburst activity, in contrast to   CzeV404~Her, which varies between 14\fm5 in a superoutburst to 17\fm5 in quiescence. The V-shaped eclipse of V348~Pup \citep{2001MNRAS.328..903R, 2010MNRAS.409.1195D} 
is similar to that of  CzeV404~Her in a superoutburst. The eclipse mapping of the accretion disk reveals two asymmetric structures in the accretion disk  \citep{2016MNRAS.457..198S}. One of them looks like a tail of the disk-stream impact region and has an azimuthal extension of about 90 deg. The other structure is more probably related to the inner part of the disk. The position of the bright region in the disk of CzeV404~Her (marked in red and blue in Fig.~\ref{Fig:10}) matches the first structure in the accretion disk of V348~Pup \citep[see their fig.~7]{2016MNRAS.457..198S}. Moreover, the Doppler maps of  CzeV404~Her are very similar to those observed in other SW~Sex-type systems \citep{1994ApJS...93..519K, 1994ApJ...431L.107H, 1997MNRAS.291..694D, 1998MNRAS.294..689H, 2013MNRAS.428.3559D, 2015gacv.workE..34S}.
As we noted before,  the accretion disk of CzeV404~Her in superoutburst is hot and flat, similar to what was reported in V Per.
Following our light-curve modelling, Doppler mapping, and the observed outburst behaviours,  we therefore propose that Czev404~Her is an example of a CV that simultaneously shows the characteristics of SU~UMa-type and SW~Sex-type objects.

\section{Conclusions}
\label{sec:conc}

We performed time-resolved photometric and spectroscopic observations of the CV CzeV404~Her that is located in the period gap. We found that the object shows outburst activity of SU~UMa-type systems.  The time between normal outbursts is about 7~days, and superoutbursts occur with an interval of about 140~days. These two timescales are relatively short compared to similar period-gap CVs.  
The orbital period has not varied during  ten~years of monitoring.
Using the light curves of eclipses in quiescence and superoutbursts, we determined the fundamental parameters of the system. The object contains the massive white dwarf primary,  $M_{\mathrm{WD}}=1.00(2)$ M$_{\sun}$, and a hot secondary with an effective temperature of $T_2 = 4100(50)$~K. Another example of a hot secondary in period-gap CVs is SDSS J0011-0647. \citet{2014ApJ...790...28R} proposed that these hot secondaries are formed by evolved stars that are not expected to become fully convective \citep{2000MNRAS.318..354B}, and they therefore should not lock magnetic braking like the donor stars in canonical CVs. For this reason, they might be visible in the period gap.
 The derived value of the mass ratio is $q = 0.16$. This is lower than the value estimated from the superhump period applying different $\epsilon\sim q$ relationships. However, we note that a similar discrepancy is also observed in some other eclipsing systems \citep{2013PASJ...65..115K}. 
The system inclination is $i=78.8^\circ$. 

Although Czev404 Her shows outburst activity characteristic of SU~UMa-type systems, the structure of its accretion disk is similar to that observed in SW~Sex-type systems: The optically thick accretion disk spreads out to the tidal limitation radius and has a  larger hot spot that extends along the outer rim of the disk.  
The spot or line is hotter than the rest of the outer part of the disk in quiescence and in the intermediate state, and it does not stand out from the disk flux in (super)outbursts. 
The eclipsing light curve, the trailed spectra of Balmer and He~I emission lines, and the Doppler maps of the object are similar to those observed in V348~Pup, another eclipsing CV located in the period gap at a similar orbital period that is classified as an SW~Sex-type system.
Based on our analyses, we therefore claim that Czev404~Her represents a link between two distinct classes of CVs: SU~UMa-  and SW~Sex-type CVs. The accretion disk structure is similar to that of SW~Sex systems. The physical conditions inside the disk begin to manifest the behaviour of SU~UMa-type objects, however.

It is generally accepted that the period-gap phenomenon in the CV period distribution is caused by a reduction in the angular momentum loss rate and a temporary switch-of off the mass transfer rate. CVs transform into detached systems with a white dwarf plus a main-sequence star that do not exhibit mass transfer activity. 
However, as we noted above, the number of CVs in the period gap has increased in the past decades, and detailed studies of these objects are essential to understand the physical mechanism of the mass transfer lock.

\medskip
\begin{acknowledgements}
The authors acknowledge allotment of observing time at San Pedro M\'artir observatory, UNAM, Baja California, Mexico.
S.Z. acknowledges PAPIIT grants IN102120.
M.W. was supported by the Czech Science Foundation grant GA19-01995S.
The research of J.K. and M.W. was supported partially by the project Progress Q47 {\sc Physics} of the Charles
University in Prague. 
This work is supported by MEYS (Czech Republic) under the projects MEYS LM2018505, LTT17006 and EU\slash MEYS
CZ.02.1.01\slash  0.0\slash 0.0\slash 16\_013\slash 0001403 and CZ.02.1.01\slash 0.0\slash 0.0\slash 18\_046\slash
0016007. The following internet-based resources were used in research for this paper:
the SIMBAD database and the VizieR service operated at CDS, Strasbourg, France;
the NASA's Astrophysics Data System Bibliographic Services.
This research is part of an ongoing collaboration between professional astronomers and the Czech Astronomical
Society, Variable Star and Exoplanet Section.

\end{acknowledgements}

\bibliography{bibliography}

\end{document}